\documentclass{aa}

\usepackage{graphicx}
\usepackage{longtable}
\usepackage{placeins}
\usepackage{txfonts}
\usepackage{natbib}
\usepackage{multirow}
\usepackage{lipsum}
\usepackage[utf8]{inputenc}
\usepackage{cite}
\usepackage[hidelinks]{hyperref}
 \hypersetup{
colorlinks   = true,
linkcolor    = red, 
citecolor   = blue 
}

\begin{document}

   \title{Photometric study of hot Algol-type binaries with long cycles}

  \subtitle{}

   \author{J. Garc\'es \inst{1} \fnmsep\thanks{juangarces@udec.cl}, R.E. Mennickent \inst{1}, J. Petrov\'ic \inst{2}, D. Barr\'ia \inst{4}, L. Celed\'on \inst{5}, C. C. Cort\'es \inst{4,6}, G. Djura{\v{s}}evi{\'c} \inst{2,3}, D.R.G. Schleicher \inst{7}, I. Soszy\'nski \inst{8}
   }
   \institute{Universidad de Concepci\'on, Departamento de Astronom\'ia, Casilla 160-C, Concepci\'on, Chile
   \and
   Astronomical Observatory, Volgina 7, 11060 Belgrade 38, Serbia
   \and
   Issac Newton institute of Chile, Yugoslavia Branch, 11060, Belgrade, Serbia
   \and
   Centro de Investigaci\'on en Ciencias del Espacio y F\'isica Te\'orica, Universidad Central de Chile, Av. Francisco de Aguirre 0405, La Serena, Chile
   \and
   Instituto de F\'isica y Astronom\'ia, Universidad de Valpara\'iso, Av. Gran Breta\~na 1111, Valpara\'iso, Chile
   \and
   Departamento de Tecnolog\'ias Industriales, Faculty of Engineering, Universidad de Talca, Merced 437, Curic\'o, Chile
   \and
   Dipartimento di Fisica, Sapienza Università di Roma, Piazzale Aldo Moro 5, 00185 Rome, Italy
   \and
   Astronomical Observatory, University of Warsaw, Al. Ujazdowskie 4, 00-478 Warszawa, Poland
   }

   \date{Received November 14, 2024; accepted July 09, 2025}

\abstract
{Double periodic variables (DPVs) are hot Algol-type interacting binary systems with an orbital and a long photometric cycle. The origin of the latter may be related to cyclic structural changes in the accretion disc that surrounds the gainer star that are driven by a variable mass-transfer rate. If this is the case, changes in the orbital light curve would be expected throughout the long cycle.}
{We conducted a detailed photometric analysis of the light curves of 134 Large Magellanic Cloud (LMC) DPVs to investigate variations in the morphology of the orbital light curves as a function of the long-cycle phase.}
{We separated the two photometric cycles from the Optical Gravitational Lensing Experiment (OGLE) {\it
I }band light curves for the systems. We thus compared the orbital light curves at opposite long-cycle phases, investigated the stability of the long period, and analysed the residuals of the separation process to search for significant frequencies above a 1\% false-alarm probability threshold.}
{We confirm that the DPVs OGLE-LMC-DPV-097 and OGLE-BLG-ECL-157529 change most strongly in their orbital light curves throughout the long cycle. By comparison, about 50\% of the sample exhibits moderate morphological variations, in particular, around orbital phase 0.5. This is likely associated with structural changes in the accretion discs. In addition, we identified 18 DPVs with variable long periods, including 10 new cases. In some of them, the long period either increases or decreases continuously over time. For the first time, we found DPV systems that alternate between the two behaviours at different epochs. Moreover, we detected frequencies in the residuals that might be directly related to changes in the morphology of the orbital curves. Finally, some previously reported frequencies disappear when a variable long period is taken into account.}


  \authorrunning{Garc\'es et al.}
\titlerunning{Photometric study of LMC DPVs}

   \maketitle

\section{Introduction}

\begin{figure*}
\includegraphics[width=1 \linewidth]{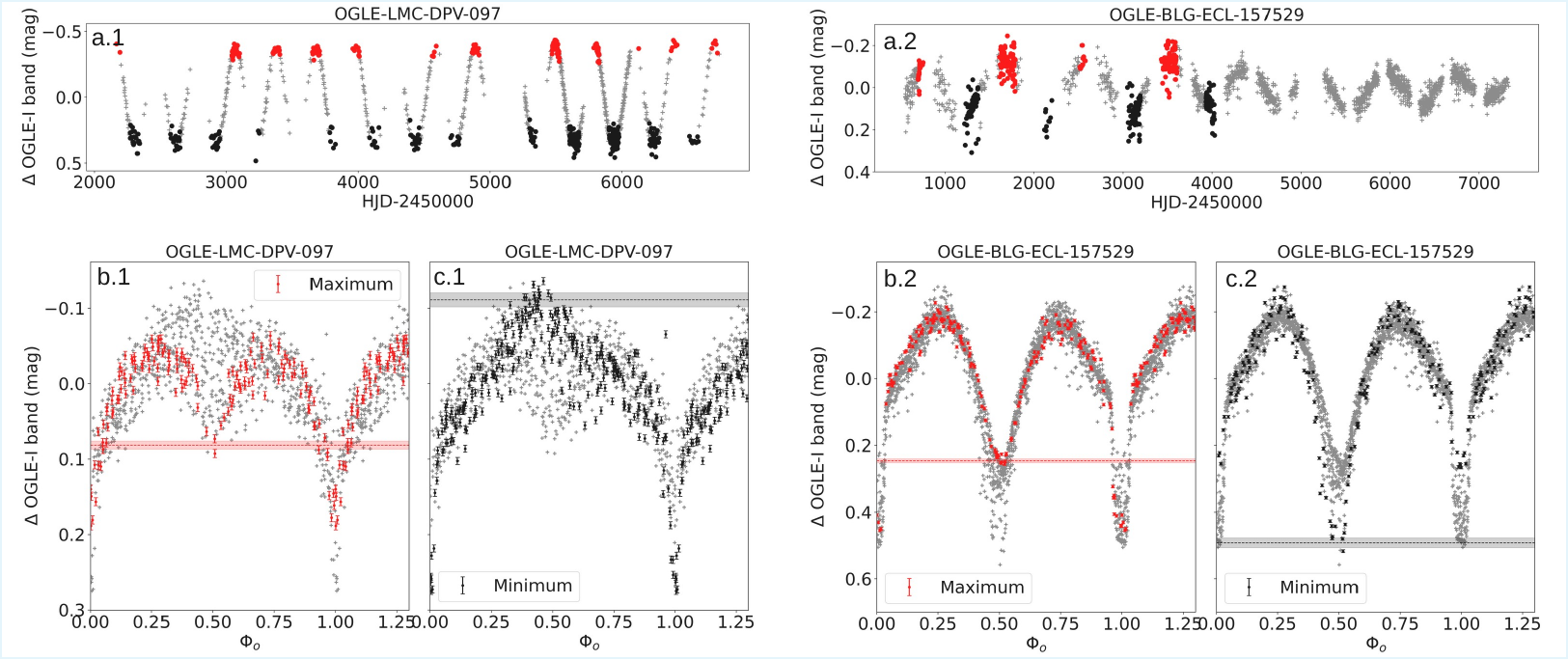}
\caption{Long cycle after the separation for OGLE-LMC-DPV-097 and OGLE-BLG-ECL-157529 (panels a.1 and a.2). The red and black dots indicate the observation epochs corresponding to the maximum and minimum of the long cycle according to the ranges defined in Sect.~\ref{long-cycle-stages}. Panels b.1, b.2, c.1, and c.2 present the orbital light curves during the maximum and minimum phases of the long cycle for the two systems. The dashed line marks the estimated magnitude near orbital phase $\Phi_o \sim 0.5$, and the shaded region represents the 1$\sigma$ confidence interval (see Sect.~\ref{lc_changes}). The grey dots in panels b.1, b.2, c.1, and c.2 correspond to the same observation epochs as in panels a.1 and a.2. For OGLE-BLG-ECL-157529, only epochs with the largest long-cycle amplitude are displayed (see panel a.2).}
\label{fig1}
\end{figure*}
Objects in the subgroup of hot Algol-type binaries that are known as double periodic variables (DPVs) exhibit two photometric cycles in their light curves, one with an orbital period ($P_o$) whose morphology is typical of eclipsing or ellipsoidal binaries, and the other with a long period ($P_l$) that is characterised by a sinusoidal or sometimes double-hump morphology whose origin is still under debate. The two periods are closely linked, with $P_{l} \sim 35.2 \times P_{o}$ \citep{Mennickent2003}. These systems consist of an evolved A-, F-, or G-type donor that has filled its Roche lobe and is transferring mass to a B-type dwarf star (the gainer) through an optically thick accretion disc \citep{Mennickent2017}. To date, over 340 DPVs have been identified. They are distributed across the Large Magellanic Cloud \citep[LMC;][]{Mennickent2003, Poleski2010, Glowacki2024}, the Small Magellanic Cloud \citep{Pawlak2013}, and the Milky Way \citep{Mennickent2016, Rojas2021}.

The physical origin of the long cycle remains an open question, but one of the most promising scenarios suggests that it may be linked to magnetic activity in the donor star. Based on Applegate’s mechanism \citep{Applegate1992}, \citet{Schleicher2017} proposed that a dynamo acting on the donor modulates its quadrupole moment, which leads to changes in the equatorial radius and, consequently, to variations in the mass-transfer rate on timescales that are comparable to the duration of the long cycle observed in DPVs. This scenario may offer a unifying framework for interpreting key observational features modulated by the long cycle. For instance, V393 Sco exhibits a bipolar wind whose strength varies with the long-cycle phase \citep{Mennickent2012b}, while chromospheric emission lines from the donor star, such as $\ion{Fe}{II}$ and $\ion{Ti}{II}$, become enhanced near the long-cycle maximum, possibly due to increased magnetic activity \citep{Mennickent2012b, Mennickent2018}. In $\ion{HD}{170582}$, Doppler tomography of the \ion{H}{$\alpha$} emission line revealed changes in the emissivity of the stream–disc impact region during the long cycle, suggesting a modulation of the mass-transfer rate \citep{Mennickent2017}. Furthermore, a high-resolution spectroscopic study of $\rm AU\,Mon$ concluded that the long cycle is driven by structural changes in the accretion disc \citep{Armeni2022}.

Long-term photometric monitoring, such as that carried out by the Optical Gravitational Lensing Experiment (OGLE) project, has also become essential for advancing the understanding of DPVs. In this context, OGLE-LMC-DPV-097 and OGLE-BLG-ECL-157529 stand out by revealing key features in their orbital light curves that vary systematically throughout the long cycle \citep{Poleski2010, Garces2018, Mennickent2020, Mennickent2021}.

OGLE-LMC-DPV-097 displays the largest sinusoidal long-cycle amplitude ($A_{LC}$ = 0.717 mag) of the DPVs in the OGLE {\it I} band. Initial reports by \citet{Poleski2010} indicated that the depths of the primary and secondary minima vary as a function of the long-cycle phase. During the maximum phase, the orbital light curve shows well-defined primary and secondary eclipses, with both quadratures reaching similar maximum brightness. In contrast, during the minimum phase, the secondary eclipse disappears entirely, while the primary becomes significantly deeper (see Fig. \ref{fig1}, left panel; shown here for illustrative purposes). These orbital changes were first modelled by \citet{Garces2018}, who demonstrated that they are consistent with structural variations in the accretion disc; it appears hotter (6870 K) and smaller (7.5 $R_{\odot}$) during the minimum phase of the long cycle and cooler (5580 K) and larger (15.4 $R_{\odot}$) during the maximum phase.

OGLE-BLG-ECL-157529 also exhibits notable changes in its orbital light curve over the long cycle. At certain epochs, a reversal of the primary and secondary minima was reported, along with evidence that the long-cycle period decreases over time \citep{Mennickent2020}. During the maximum phase of the long cycle, the primary minimum occurs at orbital phase 0 or 1; during the minimum phase, however, the primary minimum shifts to orbital phase 0.5 (see Fig. \ref{fig1}, right panel; shown here for illustrative purposes). This behaviour is likely linked to structural changes in the accretion disc, which appears larger (33.2 $R_{\odot}$), hotter (3560 K), and thicker (10.5 $R_{\odot}$) during the long-cycle minimum, but becomes smaller (27.4 $R_{\odot}$), cooler (2970 K), and considerably thinner (2.9 $R_{\odot}$) during the maximum phase \citep{Mennickent2020}. It has been proposed for this system that the observed changes are driven by a variable mass-transfer rate, and that the long cycle results from the partial occultation of the gainer by the accretion disc \citep{Mennickent2021}.

In orbital phase 0.5, the photometric behaviour of these two systems differs as a function of the long-cycle phase. In OGLE-LMC-DPV-097, the secondary minimum is deeper during the long-cycle maximum than its opposite phase. In contrast, OGLE-BLG-ECL-157529 shows the opposite trend at the same orbital phase (see Fig.~\ref{fig1}). This suggests that the orbital modulations may differ significantly in differnt DPVs. If these changes are driven by variations in the accretion disc, then the associated structural changes are also likely to vary from system to system. For this reason, we conducted a photometric analysis of 134 systems to investigate whether these cyclic variations are widespread. The identification of these photometric patterns might be key to understanding the role of accretion disc changes in driving the long cycle.

\section{Photometric data from the OGLE catalogue}
\label{photometry-data}

Our study was conducted based on OGLE photometry from the OGLE-II \citep{Udalski1997}, OGLE-III \citep{Udalski2003}, and OGLE-IV \citep{Udalski2015} databases that was obtained with the 1.3 m Warsaw telescope at Las Campanas Observatory, Chile, operated by the Carnegie Institution for Science.
The photometry was obtained using the difference image analysis (DIA) \citep{Alard1998, Alard2000, Wozniak2000}, which employs an image-subtraction algorithm and is highly efficient in dense star fields. The photometry obtained in OGLE consistently used filters close to the standard photometric $V$ and $I$ bands. This has enabled a straightforward transformation and calibration of the OGLE photometry to the standard system in its different campaigns. The OGLE {\it I} band (centred at 8000 $\AA{}$) closely resembles the standard Kron-Cousins {\it I} band filter, while the OGLE {\it V} band (centred at 5500 $\AA{}$) differs slightly from the standard Johnson $V$ band \citep{Udalski2015}.

We analysed 134 DPVs in the LMC, all of which are listed in the catalogue published by \citet{Poleski2010}. The OGLE IDs of all analysed objects are summarised in Table~\ref{tab-res-analisis-1}. Their photometric data are publicly available through the OGLE\footnote{\url{https://ogledb.astrouw.edu.pl/\~ogle/CVS/\#dpv}} online database, which also provides relevant parameters such as the orbital and long-cycle periods, the amplitudes of the two light curves, and additional remarks for each system.
The vast majority of the catalogued LMC DPVs continue to be observed in the OGLE-IV campaign, whose photometry was kindly provided by the OGLE team. This has enabled us to accumulate more than 2000 epochs of {\it I} band observations for some objects that span over 25 years of photometric data in certain cases.

\section{Data analysis}
\label{data-analysis}

\subsection{Separating the light curves}
\label{disentangling}

It is essential to separte the orbital light curves from their respective long cycles. For this purpose, we employed an algorithm based on a Fourier component amplitude analysis that is commonly used in photometric studies of DPVs \citep[see e.g.][]{Mennickent2012a}. This algorithm requires prior knowledge of the frequencies implicit in the light curves, specifically, of $f_{o}$ ($P_{o}^{-1}$) and $f_{l}$ ($P_{l}^{-1}$), which correspond to the orbital and long-cycle frequencies, respectively. These frequencies were taken from the LMC DPV catalogue \citep{Poleski2010}.

In addition, it is crucial to select an appropriate number of Fourier harmonics for the fitting process. Too many harmonics might lead to overfitting, whereas too few harmonics might fail to capture the morphological features of the light curve. To avoid these issues, we determined the optimal number of harmonics by selecting the solution that minimised the reduced $\chi^2$ of the fit. This approach allowed us to recover clean orbital and long-cycle light curves that could be analysed independently.

It is also worth noting that the stability of the adopted periods is essential for a reliable disentangling process. No changes in the orbital period were reported for the DPVs analysed in this study. A variable orbital period is not a typical feature of this class of objects. Only two of the DPV systems known to date have shown evidence of such variations: $\beta$~Lyr \citep{Harmanec1993} and RX~Cas \citep{Mennickent2022}. Some DPVs exhibit variable long-cycle durations, however. In these cases, we applied a different disentangling strategy that we describe below.

\subsubsection{Variable long period}
\label{variable-long-period}

To date, eight LMC DPVs with variable long-cycle periods have been reported in the literature (summarised in Table~\ref{tab:period-variations}). Some of these variations are significant, such as the decrease of 75 days in the LMC DPV OGLE-LMC-DPV-065 over 19 years \citep{Mennickent2019}, and the decrease of nearly 100 days in the Galactic DPV OGLE-BLG-ECL-157529 over 18.5 years \citep{Mennickent2020}.

When the long period varied, we divided the photometric data into separate time bins, each of which contained at least two complete long cycles, and determined the dominant long-cycle period in each using the generalised Lomb-Scargle (GLS) periodogram \citep{zechmesiter09} that is implemented in the \texttt{PyAstronomy} package for Python. Within each bin, the long period was assumed to remain constant. A similar method was applied to OGLE-LMC-DPV-065, where the available photometry was divided into six time bins and was disentangled individually \citep[for details see][]{Mennickent2019}.

While the GLS periodogram was used to estimate the long-cycle period in specific time bins for disentangling purposes, we complemented this approach by performing a time–period analysis of the separated long-cycle light curve using the weighted wavelet Z-transform \citep[WWZ; see][]{Foster1996}. The WWZ handles changes in frequency over time and is robust to irregular data. It is therefore an ideal technique for investigating long-term variability in binary systems. For our analysis, we adopted a decay parameter of $\alpha = 0.001$ and explored period ranges that broadly encompass the published long-cycle period of each object. In all cases, the resolution of the time–period maps was one day in the period domain and 200 bins in heliocentric Julian date (HJD).

\subsubsection{Searching for frequencies in the residuals}
\label{frequency-combinations}

\begin{table}
\centering
\caption{DPVs with reported long-period variations.}
\label{tab:period-variations}
\begin{tabular}{lc}
\hline
\hline
Object ID & Period Variation \\
\hline
OGLE-LMC-DPV-049$^{(1)}$       & Decrease \\
OGLE-LMC-DPV-052$^{(1)}$       & Decrease \\
OGLE-LMC-DPV-056$^{(2)}$       & Decrease \\
OGLE-LMC-DPV-065$^{(1,3,5)}$   & Decrease \\
OGLE-LMC-DPV-080$^{(1)}$      & Increase \\
OGLE-LMC-DPV-095$^{(1)}$       & Decrease \\
OGLE-LMC-DPV-110$^{(1)}$       & Decrease \\
OGLE-LMC-DPV-114$^{(1)}$       & Decrease \\
OGLE-BLG-ECL-157529$^{(4)}$    & Decrease \\
\hline
\end{tabular}
\tablefoot{Parenthetical numbers refer to the references (1)~\citet{Mennickent2005b}, (2)~\citet{Mennickent2008}, (3)~\citet{Mennickent2019}, (4)~\citet{Mennickent2020}, and (5)~\citet{Poleski2010}.}
\end{table}

After removing the orbital and long-cycle frequencies from the raw light curves through the disentangling process, we analysed the residuals. Specifically, we studied two types of frequencies that were reported previously.
The first type corresponds to linear combinations of the orbital ($f_o$) and long-cycle ($f_l$) frequencies. The LMC DPV catalogue presented by \citet{Poleski2010} reported these frequency combinations for 36 systems, with the most common being $f_c = f_o + f_l$, which was observed in 30 systems. These combinations were also confirmed in 7 of the 10 cases studied by \citet{Buchler2009}.
The second type includes additional frequencies that are not associated with linear combinations of $f_o$ and $f_l$. These were identified in 21 systems and are also documented in the remarks section of the OGLE LMC DPV catalogue \citep{Poleski2010}.
To search for these frequencies, we applied the GLS periodogram to the residuals and considered all peaks that exceeded the 1\% false alarm probability (FAP) threshold. For each significant detection, the GLS analysis yielded the amplitude of the associated variability along with its uncertainty.

\subsection{Ephemerides of the light curves}

We defined the phase of each observation using linear ephemerides of the form $T = T_0 + P \times E$, where $T_0$ is the reference epoch, $P$ the period, and $E$ the number of elapsed cycles. For the orbital and long-cycle light curves, we adopted the ephemerides provided in the OGLE LMC DPV catalogue \citep{Poleski2010}. These ephemerides are reported with high precision, including uncertainties in the orbital and long periods (e.g. $P_o = 7^{d}.751749 \pm 0.000202$ and $P_{l} = 302^{d}.622 \pm 0.109$ for OGLE-LMC-DPV-097). This ensures reliable phase calculations. In the case of the orbital light curves, $T_0$ corresponds to the epoch of the primary minimum, while for the long-cycle light curves, $T_0$ refers to the epoch of maximum brightness. In systems with a variable long period, we divided the photometric data into separate time bins, as described in Sect.~\ref{variable-long-period}, and computed local ephemerides for each. This procedure allowed us to construct a single light curve using all the available photometry.

This approach ensured that the orbital and long-cycle phases were defined consistently across the dataset. This allowed us an unambiguous comparison of the orbital light curves in different long-cycle stages.

\subsection{Long-cycle stages}
\label{long-cycle-stages}

Since in most LMC DPVs the long cycle have a sinusoidal morphology, we divided this cycle into two key phases: maximum ($\rm 0.9 \leq \rm \Phi_{l} < \rm 1.1$) and minimum ($\rm 0.4 \leq \rm \Phi_{l} < \rm 0.6$). When the long cycle showed a double-hump morphology, as seen in OGLE-LMC-DPV-056 \citep{Mennickent2008}, we used different ranges. Although in these cases, the main maximum of the long cycle also occurs at $\rm 0.9 \leq \Phi_{l} < \rm 1.1$, we chose two ranges for the minimum ($\rm 0.2 \leq \Phi_{l} < \rm 0.4$ and $\rm 0.6 \leq \Phi_{l} < \rm 0.8$), as was done in the case of OGLE-LMC-DPV-065 \citep{Mennickent2019}. We then assigned each observation of the orbital cycle to the corresponding long-cycle phase using the relevant ephemerides. This procedure enabled us to construct orbital light curves in the maximum and minimum long-cycle stages, based on which, we directly compared their morphology, as was done in the case of OGLE-LMC-DPV-097 \citep{Garces2018}.

\subsection{Changes in the orbital light curve}
\label{lc_changes}

\begin{figure*}[h]
\centering
\begin{tabular}{l l l}
\includegraphics[width=0.315\linewidth]{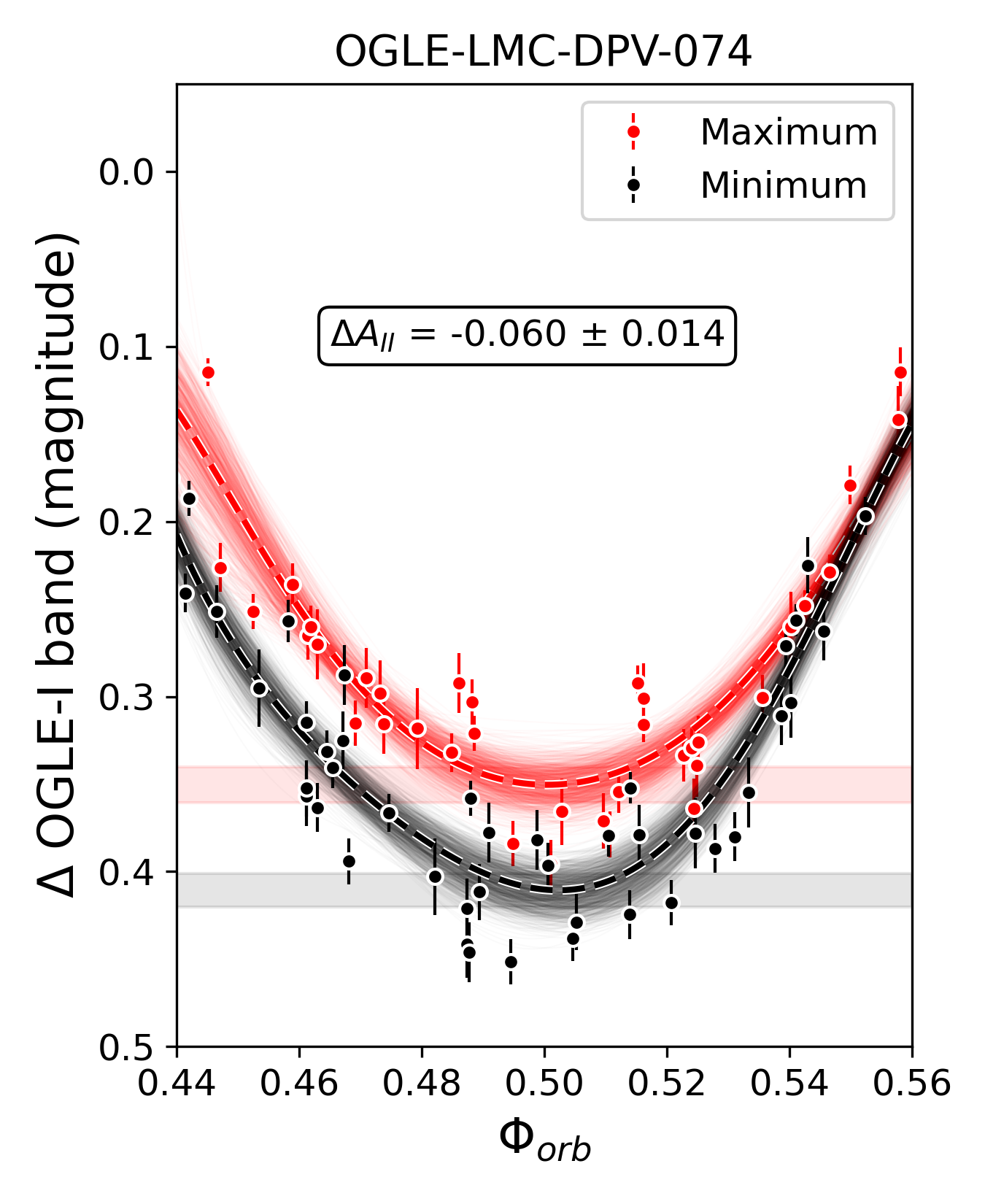} &
\includegraphics[width=0.315\linewidth]{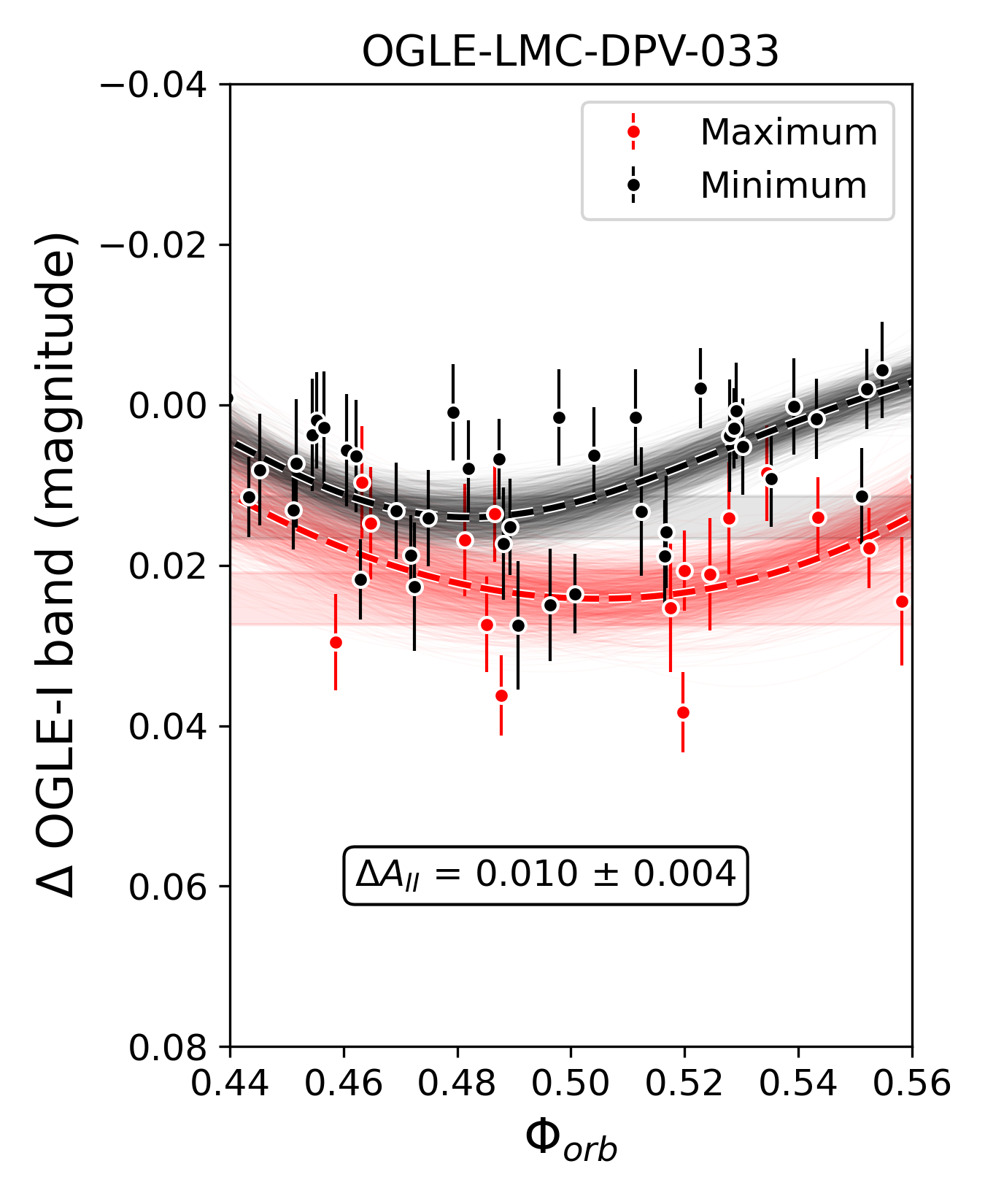} &
\includegraphics[width=0.315\linewidth]{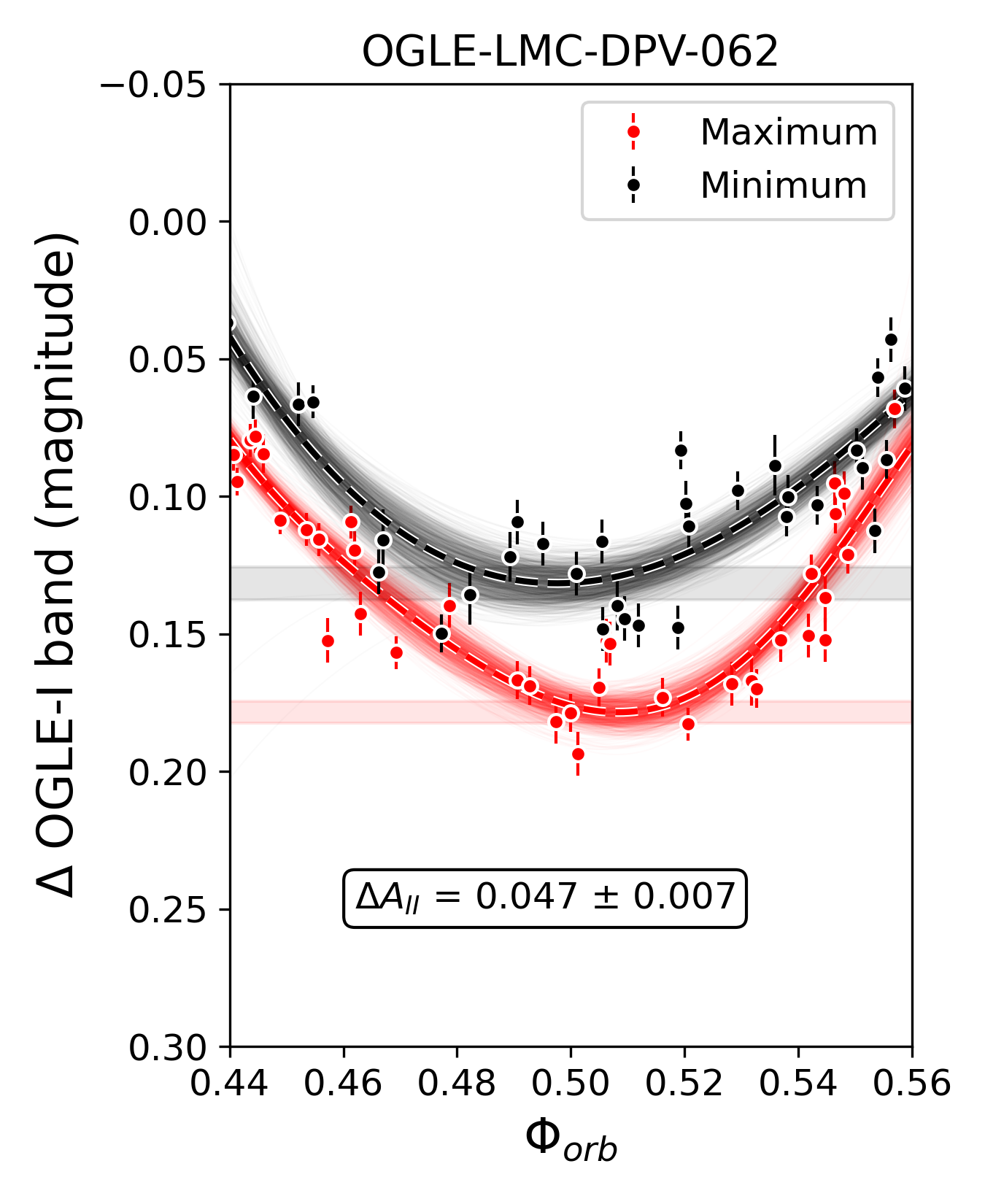} \\
\end{tabular}
\caption{Photometry of the orbital cycle around orbital phase 0.5 for three DPVs: OGLE-LMC-DPV-074 (left panel), OGLE-LMC-DPV-033 (middle panel), and OGLE-LMC-DPV-062 (right panel). The dashed red and black lines represent the best polynomial fits to the data at the maximum and minimum phases of the long cycle, respectively. The solid lines correspond to 1000 bootstrap realisations that we used to estimate the uncertainty, and the shaded region indicates the resulting 1$\sigma$ confidence interval (see details in Sect.~\ref{lc_changes}). The quantity $\Delta A_{II}$ is derived using Eq.~\ref{eq-amp}.}
\label{fig-cambios-minII}
\end{figure*}

Theoretical models applied to the orbital light curves of various DPVs consider a binary configuration consisting of a massive star surrounded by an accretion disc and an evolved star that has filled its Roche lobe. These models suggest that at the primary minimum ($\Phi_{o} = 0.0$), the donor star eclipses the gainer star and the disc, whereas at the secondary minimum ($\Phi_{o} = 0.5$), the gainer star and the disc eclipse the donor \citep[for a detailed description see e.g.][]{Mennickent2012b, Garrido2013, Garces2018, Mennickent2021}.

Changes in the morphology of the orbital light curve throughout the long cycle, in particular, around orbital phase 0.5, are clearly observed in systems such as OGLE-LMC-DPV-097 and OGLE-BLG-ECL-157529 (see Fig.~\ref{fig1}). In both cases, these variations have been attributed to structural changes in the accretion disc \citep{Garces2018, Mennickent2021}. For this reason, it is essential to examine the photometric behaviour around this orbital phase over the long cycle.

To quantify the observed variability, we defined the amplitude, $A_{II}$, as the difference between the magnitude at orbital phase $\Phi_o \sim 0.5$ and at $\Phi_o \sim 0.25$, corresponding to the secondary minimum and the first quadrature maximum, respectively (i.e. $A_{II} = \mathrm{Mag}_{0.5} - \mathrm{Mag}_{0.25}$). The magnitudes around $\Phi_o \sim 0.5$ and $\Phi_o \sim 0.25$ were estimated by fitting a polynomial to the data restricted to the phase ranges $0.4$–$0.6$ and $0.1$–$0.4$, respectively. The fits were performed using the \texttt{numpy.polyfit} function in \texttt{Python}, weighted by the photometric errors and evaluating polynomial degrees from 3 to 7. The optimal degree was selected by minimising the reduced $\chi^2$. To estimate the statistical uncertainty, we applied the bootstrap resampling method described by \citet{Efron1982}. We generated 1000 random samples with replacement from the original data and fitted each sample using the same optimal polynomial order. The final uncertainty was estimated as the standard deviation ($1\sigma$) of the resulting distribution, while the reported minimum or maximum corresponds to the value obtained from the best fit.

We therefore defined
\begin{equation}
\label{eq-amp}
\Delta A_{II} = A_{II, \max} - A_{II, \min},
\end{equation}
where $A_{II, \max}$ refers to the amplitude of the orbital light curve near the secondary minimum during the maximum phase of the long cycle, and $A_{II, \min}$ corresponds to the same measurement during the long-cycle minimum. Since the orbital light curves were separated from the long-cycle variability and are expressed in differential magnitudes, the brightness near the first quadrature ($\Phi_o \sim 0.25$) remained approximately constant between opposite long-cycle phases. As a result, $\Delta A_{II}$ primarily reflects the change in brightness around $\Phi_o \sim 0.5$. It is worth noting that OGLE-LMC-DPV-097 constitutes an exception, as the maximum of the orbital light curve does not occur near $\Phi_o \sim 0.25$, but rather close to $\Phi_o \sim 0.5$ during the long-cycle minimum. To address this particular case, we adopted the value at $\Phi_o \sim 0.25$ estimated during the long-cycle maximum to compute both $A_{II, \max}$ and $A_{II, \min}$.

We then compared $\Delta A_{II}$ with the amplitude of the long cycle estimated using the GLS analysis applied to the disentangled light curve. This periodogram identifies the most significant peak associated with the long-period signal, and through the corresponding sinusoidal fit, it yields its amplitude and associated uncertainty.

This analysis is applicable to DPVs with either eclipsing or ellipsoidal orbital light curves. In ellipsoidal systems, the orbital inclination is lower than in eclipsing systems, so that the light curve is primarily modulated by the rotation of the evolved star that has filled its Roche lobe. Maxima occur near orbital phases 0.25 and 0.75, when the largest surface area of the lobe is projected along the line of sight. Although less pronounced, the primary and secondary minima can still be distinguished.

\section{Results}
\label{lc-results}

\subsection{Changes in the orbital light curves related to the long cycle}
\label{cambios-curvas}

Using a $3\sigma$ significance threshold, we noted that the mininum depth of the secondary in 50\% of the analysed DPVs varied significantly. These significant cases were divided into two subgroups: The first group, with $\Delta A_{II} > 3\sigma$, contained 51 systems (38\% of the sample) that are characterised by a deeper secondary minimum during the long-cycle maximum, similar to OGLE-LMC-DPV-097. The second group, with $\Delta A_{II} < -3\sigma$, included 16 DPVs (12\%) with the opposite behaviour, that is, a deeper secondary minimum during the long cycle minimum, as observed in OGLE-BLG-ECL-157529. On the other hand, the group without significant variations ($|\Delta A_{II}| \leq 3\sigma$) contained 67 systems (50\%), 62 of which are ellipsoidal DPVs (see Table~\ref{tab:amp-summary} for details).

In Fig.~\ref{fig-cambios-minII} we present representative examples of each group. OGLE-LMC-DPV-074 (left panel, $\Delta A_{II} = -0.060 \pm 0.014$ mag) and OGLE-LMC-DPV-062 (right panel, $\Delta A_{II} = 0.047 \pm 0.007$ mag) show significant variability and behave opposite in their secondary minimum depth as a function of the long-cycle phase. OGLE-LMC-DPV-033 (middle panel, $\Delta A_{II} = 0.010 \pm 0.004$ mag) is an ellipsoidal DPV, on the other hand, and varies insignificantly.

\begin{figure}
\centering
\begin{tabular}{c c}
\includegraphics[width=1.0\linewidth]{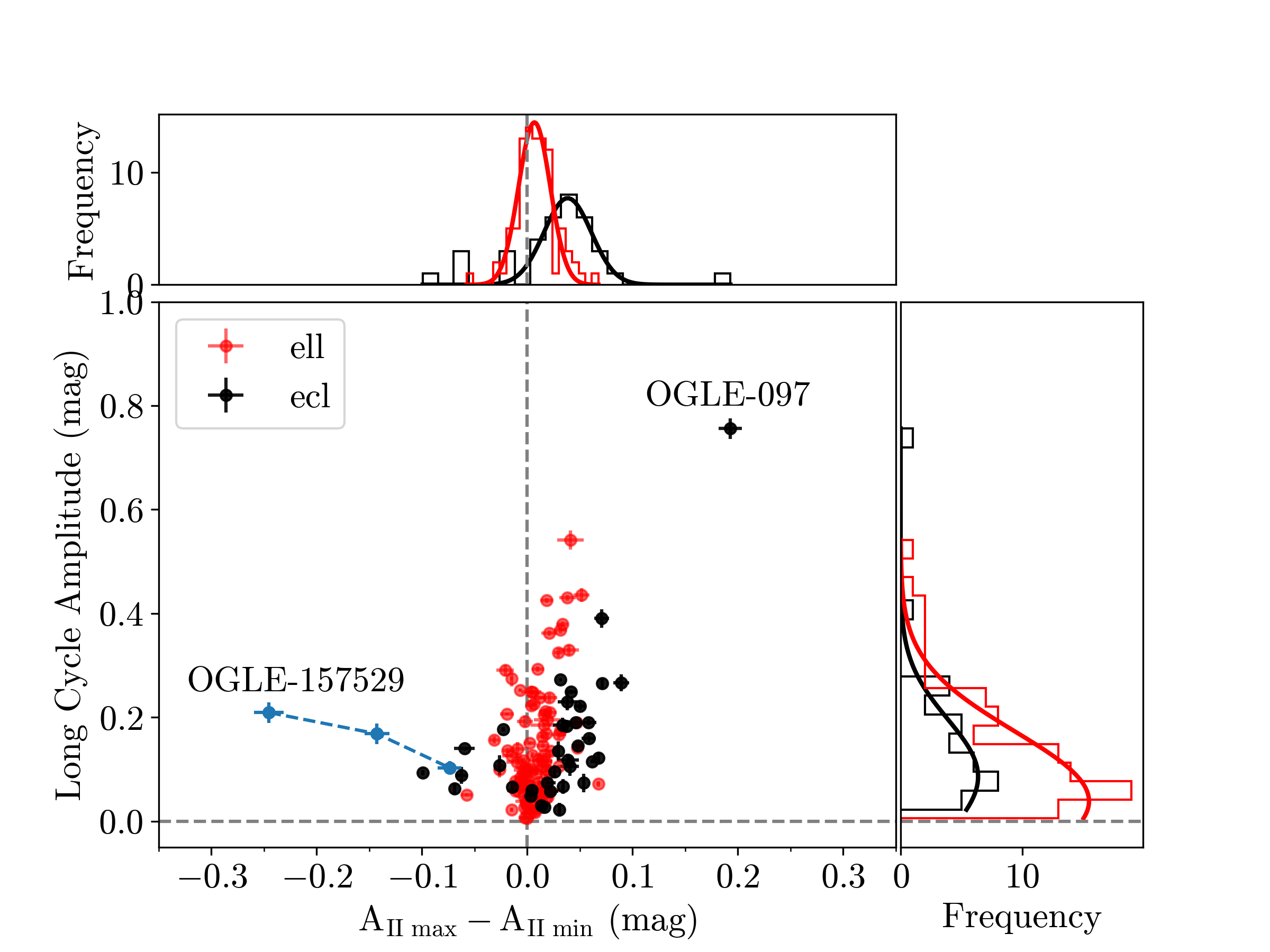}
\end{tabular}
\caption{$\Delta A_{II}$ vs. long-cycle amplitude in the OGLE {\it I} band.
The red and black dots represent binary systems classified as ellipsoidal and eclipsing, respectively. In addition, we show the location of OGLE-LMC-DPV-097 and OGLE-BLG-ECL-157529, as they represent extreme cases of changes in the morphology of the light curve. The dashed blue line traces the case of OGLE-BLG-ECL-157529, whose decrease in the long-cycle amplitude is associated with a reduced $\Delta A_{II}$.}
\label{delta-mag}
\end{figure}

A global overview of these variations is shown in Fig.~\ref{delta-mag}, which summarises the estimates of $\Delta A_{II}$ as a function of the long-cycle amplitude. As expected, OGLE-LMC-DPV-097 and OGLE-BLG-ECL-157529 vary most strongly. For OGLE-LMC-DPV-097, the variation between opposite phases of the long cycle reaches $0.193 \pm 0.011$ mag. Similarly, the variation in OGLE-BLG-ECL-157529 of $\Delta A_{II}$ decreases from $-0.245 \pm 0.014$ mag at the start of the OGLE observations to $-0.073 \pm 0.011$ mag towards the end. As these variations weaken, the amplitude of the long cycle also decreases, from $0.21 \pm 0.02$ mag to $0.10 \pm 0.01$ mag in the OGLE {\it I} band. A summary of $\Delta A_{II}$ and long-cycle amplitude for all DPVs we analysed is presented in Table~\ref{tab-res-analisis-1}.

\begin{figure*}[h]
\centering
\begin{tabular}{l l l}
\includegraphics[width=0.315\linewidth]{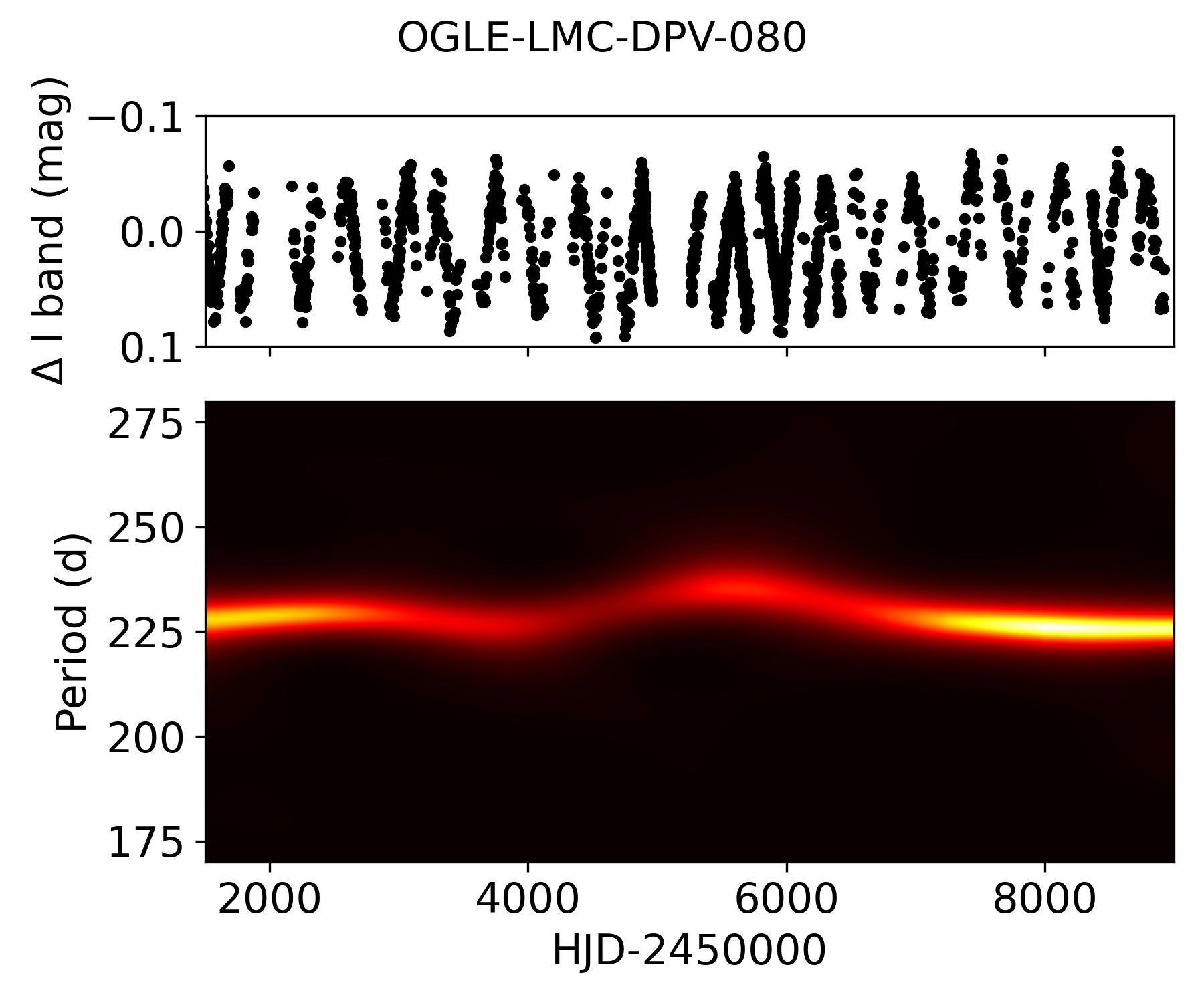} &
\includegraphics[width=0.315\linewidth]{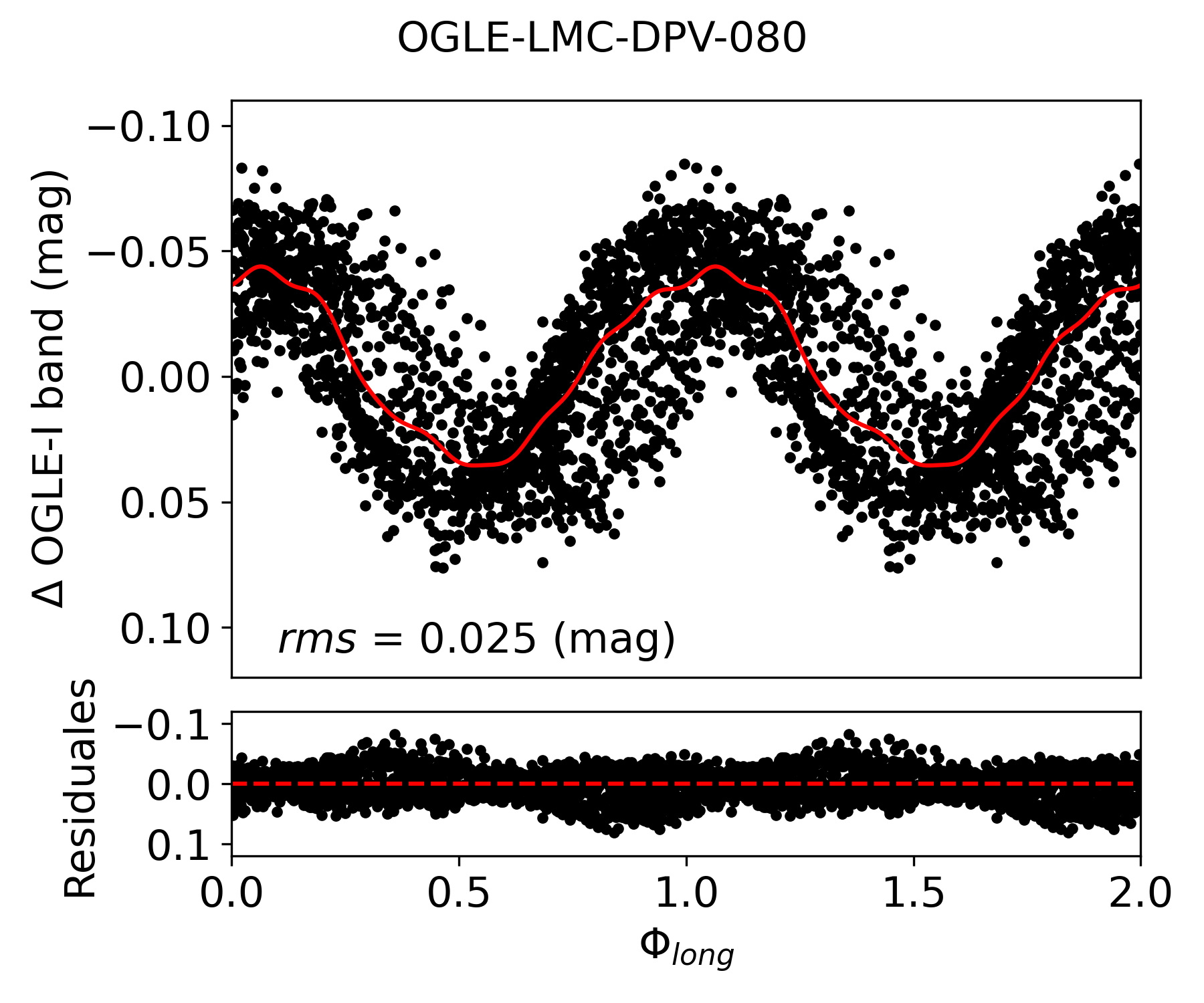} &
\includegraphics[width=0.315\linewidth]{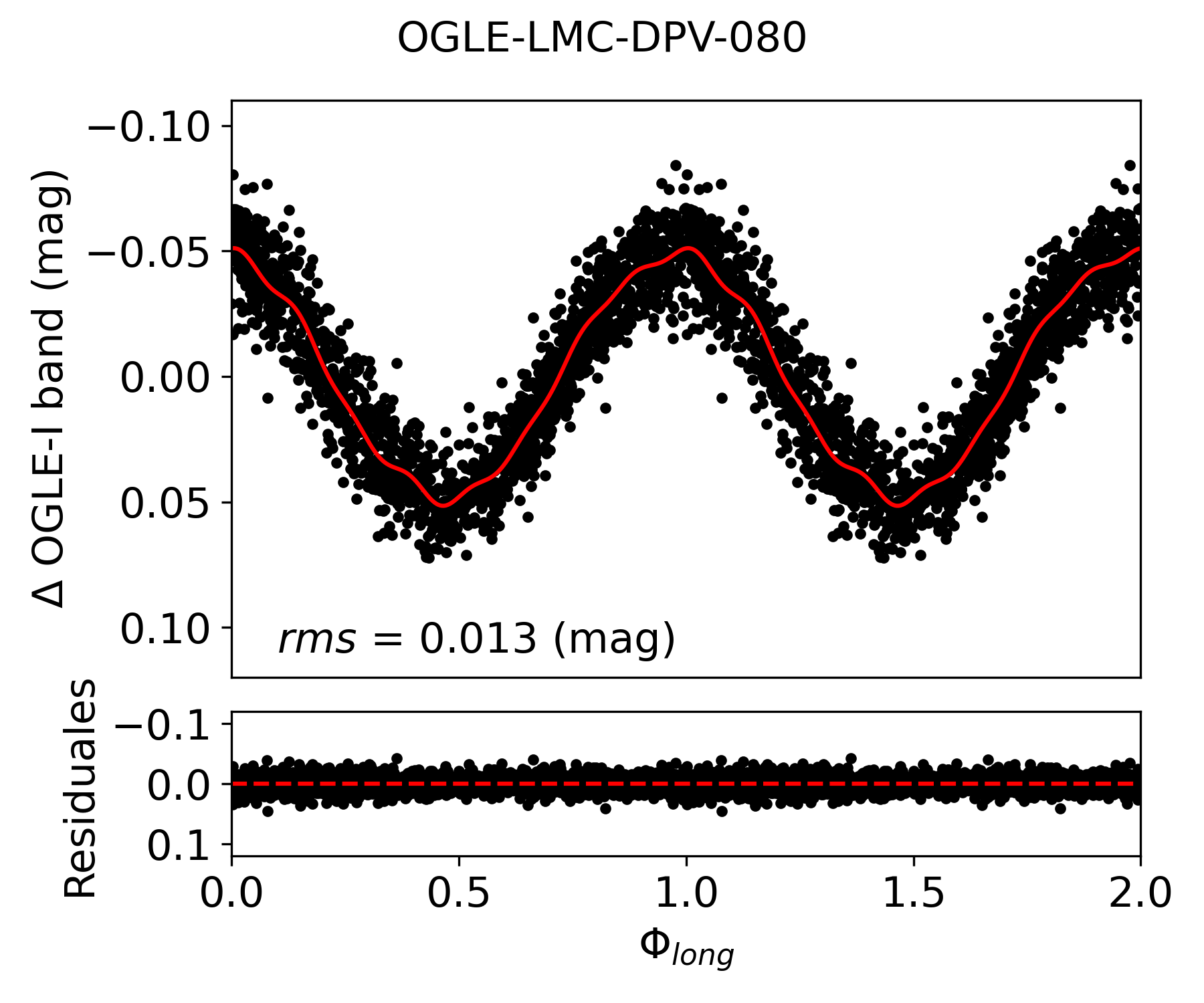} \\
\end{tabular}
\caption{Left panel: Photometry of the long cycle of OGLE-LMC-DPV-080 after the orbital cycle was removed. The WWZ analysis is shown in the lower left panel shows the evolution in the long period during the observation years. Middle panel: Long-cycle light curve assuming a constant period of 227 days. Right panel: Same as the middle panel, but considering a variable long period during the disentangling process. The dispersion of the fit to the light curve decreases considerably in the latter case.}
\label{fig-080}
\end{figure*}

\begin{table}
\centering
\caption{Median values of $\Delta A_{II}$, long-cycle amplitude $A_{LC}$, and Pearson correlation coefficient $r$, for eclipsing and ellipsoidal DPVs.}
\small
\begin{tabular}{lccc}
\hline
\hline
Group & Median $\Delta A_{II}$ & Median $A_{LC}$& $r$ \\
 & (mag) & (mag) &  \\

\hline
\multicolumn{4}{l}{$\Delta A_{II} > 3\sigma$} \\
Eclipsing (24, 18\%) & 0.044 & 0.171 & 0.86 \\
Ellipsoidal (27, 20\%) & 0.022 & 0.167 & 0.42 \\
\hline
\multicolumn{4}{l}{$\Delta A_{II} < -3\sigma$} \\
Eclipsing (7, 5\%) & -0.060 & 0.094 & --- \\
Ellipsoidal (9, 7\%) & -0.015 & 0.099 & --- \\
\hline
\multicolumn{4}{l}{$|\Delta A_{II}| \leq 3\sigma$} \\
Eclipsing (5, 4\%) & 0.013 & 0.059 & --- \\
Ellipsoidal (62, 46\%) & 0.005 & 0.084 & --- \\
\hline
\end{tabular}
\tablefoot{The number of systems and their percentage relative to the full sample (134 DPVs) are indicated in parentheses. Groups are defined using the 3$\sigma$ significance threshold.}
\label{tab:amp-summary}
\end{table}

\begin{table*}
\centering
\caption{Summary of the long-period analysis for LMC DPVs with variable long-cycle durations.}
\label{tab-cam-per}
\begin{tabular}{l c c c c c c c}
\hline
\hline
OGLE-ID       & Behaviour  & Bins & $Pl_{min}$ (d)   &        $Pl_{max}$ (d)  &      $ rms_{1}$ (mag) &         $ rms_{2} $ (mag)  & $rms$ red. (\%)\\
\hline
OGLE-LMC-DPV-010$^{ \rm New}$  & Mixed & 16 &     105.22 $\pm$  1.64  &     123.33 $\pm$ 1.34     &     0.028     &       0.020   & 28$\%$ \\
OGLE-LMC-DPV-033$^{ \rm New}$ & Decrease & 9 &     370.22 $\pm$ 2.75  &     386.41 $\pm$ 3.82     &     0.028     &       0.013   & 55$\%$ \\
OGLE-LMC-DPV-049 & Increase & 6  &     355.85 $\pm$ 4.53  &     381.48 $\pm$ 4.26                           &     0.040     &       0.030   & 26$\%$ \\
OGLE-LMC-DPV-052   & Mixed & 12 &     179.33 $\pm$ 1.34   &     190.70 $\pm$ 1.25                &     0.027     &       0.023   & 15$\%$ \\
OGLE-LMC-DPV-056   & Mixed & 8 &   171.15 $\pm$ 1.75     &     186.23 $\pm$ 1.90                 &     0.045     &       0.019   & 57$\%$ \\
OGLE-LMC-DPV-065  & Decrease & 8  &   220.12 $\pm$ 2.73     &     270.29 $\pm$ 2.02              &     0.073     &       0.028   & 62$\%$ \\
OGLE-LMC-DPV-067$^{ \rm New}$ & Mixed & 15 &     177.16 $\pm$ 0.44  &     182.07 $\pm$ 0.81       &     0.020     &       0.013   & 33$\%$ \\
OGLE-LMC-DPV-080  & Mixed & 13 &     225.34 $\pm$ 1.51  &     239.31 $\pm$ 1.43                  &     0.025     &       0.013   & 48$\%$ \\
OGLE-LMC-DPV-089$^{ \rm New}$ & Mixed & 12 &     251.79 $\pm$ 1.22  &     261.19 $\pm$ 2.90       &     0.023     &       0.015   & 36$\%$ \\
OGLE-LMC-DPV-090$^{~\rm New}$  & Mixed & 15 &     177.70 $\pm$ 0.86   &     182.17 $\pm$ 0.32               &     0.033     &       0.023   & 30$\%$ \\
OGLE-LMC-DPV-095  & Decrease  & 12 &     211.42 $\pm$ 1.42   &     254.63 $\pm$ 1.51             &     0.020     &       0.012   & 42$\%$ \\
OGLE-LMC-DPV-099$^{ \rm New}$ & Mixed & 10 &     183.08 $\pm$ 1.42   &     200.72 $\pm$ 1.78      &     0.020     &       0.012   & 39$\%$ \\
OGLE-LMC-DPV-106$^{ \rm New}$ & Increase & 8  &     402.51 $\pm$ 1.96  &     416.33 $\pm$ 2.91    &     0.051     &       0.022   & 57$\%$ \\
OGLE-LMC-DPV-108$^{ \rm New}$ & Decrease & 12 &     182.85 $\pm$ 2.53   &     198.94 $\pm$ 2.33   &     0.048     &       0.019   & 61$\%$ \\
OGLE-LMC-DPV-110  & Decrease & 11 &     202.81 $\pm$ 1.87  &     215.71 $\pm$ 2.32               &     0.022     &       0.015   & 32$\%$ \\
OGLE-LMC-DPV-114          & Mixed   & 12 &     236.19 $\pm$ 1.70  &     245.23 $\pm$ 1.30                   &     0.041     &       0.023   & 41$\%$\\
OGLE-LMC-DPV-128$^{ \rm New}$ & Mixed & 12 &     142.17 $\pm$ 2.02  &     175.47 $\pm$ 1.13       &     0.029     &       0.018   & 36$\%$ \\
OGLE-LMC-DPV-135$^{ \rm New}$ & Mixed & 13 &     180.06 $\pm$ 2.07  &     198.89 $\pm$ 1.42       &     0.032     &       0.025   & 21$\%$ \\
\hline
\end{tabular}
\tablefoot{We list the OGLE ID, the long-period behaviour over time, the time bins used for the disentangling process, and the range of long periods adopted (minimum and maximum). We also provide the $rms$ of the fit to the long-cycle light curve before ($rms_1$) and after ($rms_2$) accounting for period variability, along with the percentage reduction in $rms$ ($rms$ red.). Objects labelled 'New' represent first-time detections of long-period variability.}
\end{table*}

We also analysed the linear correlation between $\Delta A_{II}$ and the long-cycle amplitude for ellipsoidal and eclipsing DPVs with statistically significant variability ($\Delta A_{II} > 3\sigma$). These are 38\% of the sample. We quantified the correlation using the Pearson correlation coefficient ($r$), which is a statistical measure of the linear dependence between two variables. For eclipsing DPVs, we obtained $r_{\mathrm{ecl}} = 0.86$ (or $0.52$ without OGLE-LMC-DPV-097), and the coefficient for ellipsoidal DPVs is $r_{\mathrm{ell}} = 0.42$. These results indicate a strong positive correlation in eclipsing DPVs and a moderate correlation in ellipsoidal systems.

A more detailed comparison between the two groups showed that the median $\Delta A_{II}$ is larger for eclipsing DPVs, with a value of 0.044~mag (0.041~mag excluding OGLE-LMC-DPV-097) compared to 0.022~mag for ellipsoidal DPVs. The long-cycle amplitudes for both groups are similar, however, with values of 0.171~mag (0.160~mag without OGLE-LMC-DPV-097) for eclipsing DPVs and 0.167~mag for ellipsoidal DPVs. The median values of $\Delta A_{II}$ and the long-cycle amplitude for all groups defined by the $3\sigma$ criterion are summarised in Table~\ref{tab:amp-summary}.

\subsection{Variable long-cycle period}
\label{variable-long-period-sec}

Our analysis revealed 18 DPVs with temporal variations in the long cycle. Eight of these were reported before (see Table~\ref{tab:period-variations}), and we confirm their variability. The remaining 10 correspond to newly identified cases. Two of these 18 systems show a sustained increase in the long period, and the long period decreases for 5 of them. In addition, the WWZ analysis revealed a more complex behaviour in 11 systems, in which the long period fluctuates without a clear trend. This type of evolution has not been reported before for DPVs. We refer to it as mixed behaviour (see Table~\ref{tab-cam-per}). The results of the time–period analysis for systems with increasing, decreasing, and mixed behaviour are presented in \href{https://doi.org/10.5281/zenodo.15881667}{Appendix B} (see the left panel of \href{https://doi.org/10.5281/zenodo.15881667}{Fig.~B.1}).

Despite variations in the long period over time, the morphology and amplitude of the associated light curve remain sufficiently stable for us to construct a single long-cycle light curve by applying an appropriate ephemeris to each time bin before we combined the data. As detailed in Sect.~\ref{variable-long-period}, this disentangling process ensured that the resulting light curve accurately reflected the long-cycle variability, even in systems with varying periods.

As an example, we show the case of OGLE-LMC-DPV-080. This system was the only DPV that was reported to exhibit an increasing long period. In our analysis, however, which includes more than 16 years of OGLE photometry, the WWZ map oscillates irregularly between approximately 225 and 239 days, without a defined long-term trend (see Fig.~\ref{fig-080}, left panel). When the variation in the long period during the disentangling process is taken into account, the resulting residual root mean square (rms) is lower by approximately 48\, \%  than when the period is assumed to be constant (see Fig.~\ref{fig-080}, middle and right panels). This significant reduction in the rms is consistently observed for all 18 DPVs with variable long periods. The long-cycle light curves before and after the variable-period correction was applied are shown in  \href{https://doi.org/10.5281/zenodo.15881667}{Appendix B} (see the central and right panels in \href{https://doi.org/10.5281/zenodo.15881667}{Fig.~B.1}).

Table~\ref{tab-cam-per} summarises the systems we analysed for long-period variability, including the number of time bins we used in the disentangling process and the corresponding minimum and maximum long-period values we obtained. It also lists the residual rms of the long-cycle light curve with and without the variable long-period approach. In some cases, the rms reduction is modest (as low as 15\,\%), but in others, it reaches up to 62\,\%.

\subsection{Frequency combination in the residuals}
\label{freq-in-res}

We identified 73 LMC DPVs with linear frequency combinations in their residuals after we removed the orbital and long cycle from the OGLE photometry. Thirty-seven of these are new detections, which brings the total to 53\% of the LMC DPV sample. Frequency combinations were reported in only 30\% of the systems before \citep{Poleski2010}. The most common combination is the linear relation $f_c = f_o + f_l$, which we identified in 53 LMC DPVs (see Table~\ref{tab-fre-com}). As examples, Fig.~\ref{Per-cambios} shows GLS periodograms for OGLE-LMC-DPV-097 and OGLE-BLG-ECL-157529 that highlight $f_c$ in the residuals.

\begin{figure}
\centering
\begin{tabular}{c c}
\includegraphics[width=1.0\linewidth]{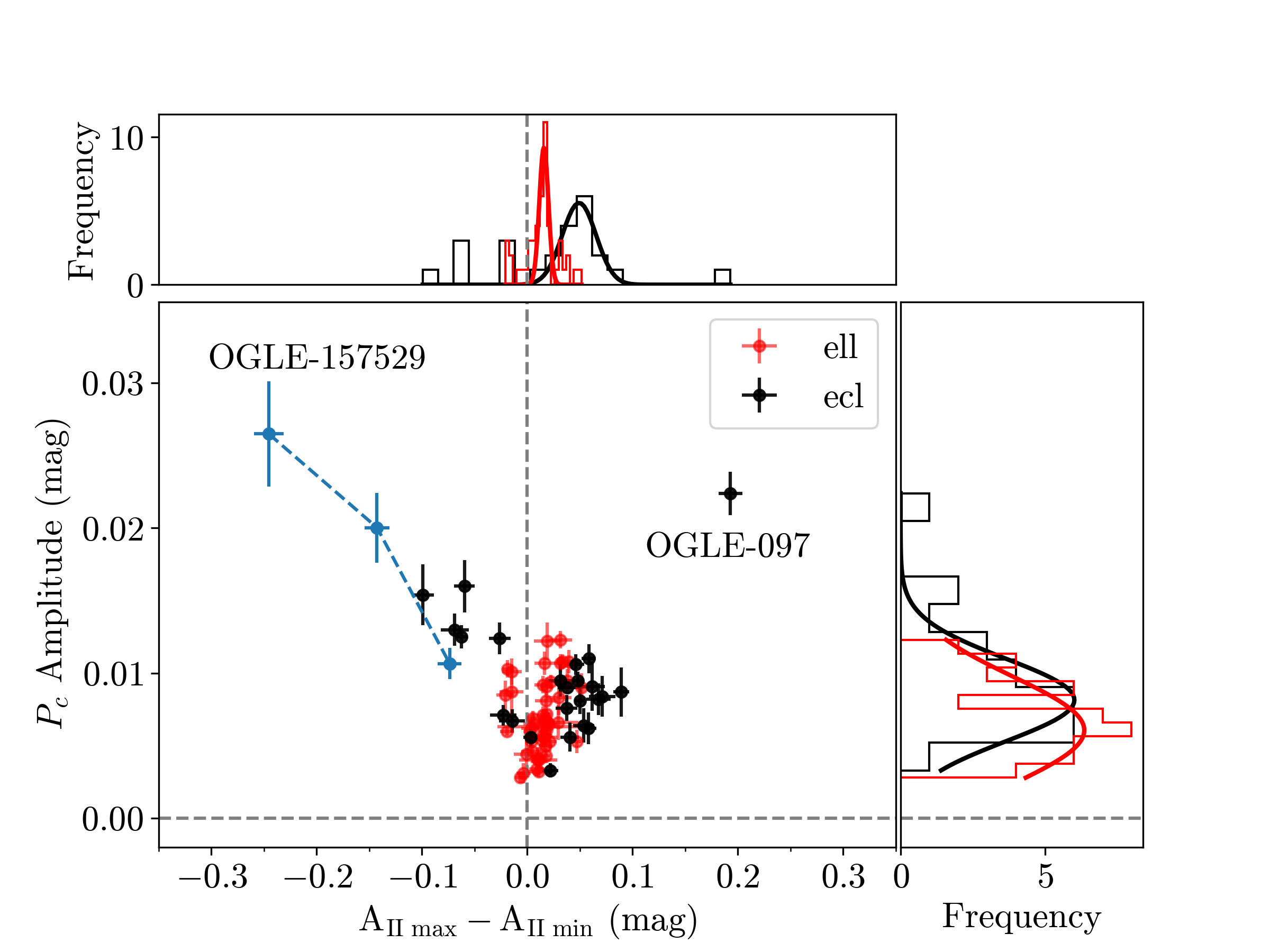}
\end{tabular}
\caption{$\Delta \mathrm{Min}_{II}$ vs. $P_{c}$ amplitude. The red and black dots represent binary systems classified as ellipsoidal and eclipsing, respectively. The positions of OGLE-LMC-DPV-097 and OGLE-BLG-ECL-157529 (blue points) stand out because their light curves have the largest amplitude in the residuals with period $P_{c}$. For OGLE-BLG-ECL-157529, the dashed blue line traces the observed trend: The amplitude of $P_{c}$ decreases as $\Delta A_{II}$ decreases.}
\label{delta-mag-GLS}
\end{figure}

Each detected frequency combination $f_c$ corresponds to a light curve with a period $P_c = 1/f_c$. We estimated the amplitude of this light curve in the residuals for all cases with a significant frequency combination. The systems OGLE-LMC-DPV-097 and OGLE-BLG-ECL-157529 show the highest $P_c$ amplitudes in our sample (see Fig.~\ref{delta-mag-GLS}). These objects have the strongest photometric variation near orbital phase $\Phi_o \sim 0.5$. For OGLE-BLG-ECL-157529, we divided the OGLE data into time bins and observed a progressive decrease in the amplitude of the $P_c$ modulation. This is accompanied by a reduction in the duration of the long-cycle period, a decrease in its photometric amplitude, and weaker changes in the orbital light curve around the secondary minimum. This correlation is clearly observed for this system alone, which provides compelling evidence that $\Delta A_{II}$ and the amplitude of the $P_c$ modulation are linked.

Figure~\ref{delta-mag-GLS} summarises the $P_c$ amplitudes for all systems with significant frequency combinations in their residuals. For DPVs in the range $0.0 \leq \Delta A_{II} \leq 0.25$, eclipsing systems exhibit a higher median $\Delta A_{II}$ (0.050~mag) than ellipsoidal systems (0.017~mag), while the median $P_c$ amplitudes are relatively similar (0.016~mag vs. 0.012~mag, respectively). The Pearson correlation coefficient between $\Delta A_{II}$ and $P_c$ amplitude is $r_{\mathrm{ecl}} = 0.88$ for eclipsing systems (or $0.44$ without OGLE-LMC-DPV-097) and $r_{\mathrm{ell}} = 0.53$ for ellipsoidal systems. For eclipsing DPVs with $\Delta A_{II}$ in the range $-0.25 \leq \Delta A_{II} \leq 0.0$, including OGLE-BLG-ECL-157529, the linear correlation is strongly negative, $r_{\mathrm{ecl}} = -0.95$. These results suggest a link between orbital light curve changes and $P_c$ amplitude, in particular, in eclipsing systems. Table~\ref{tab-fre-com} lists all frequency combinations we identified and the amplitude of their associated $P_c$ curves, including those that were reported previously by \citet{Buchler2009} and \citet{Poleski2010}.

\begin{figure}
\begin{center}
\includegraphics[width=0.99\linewidth]{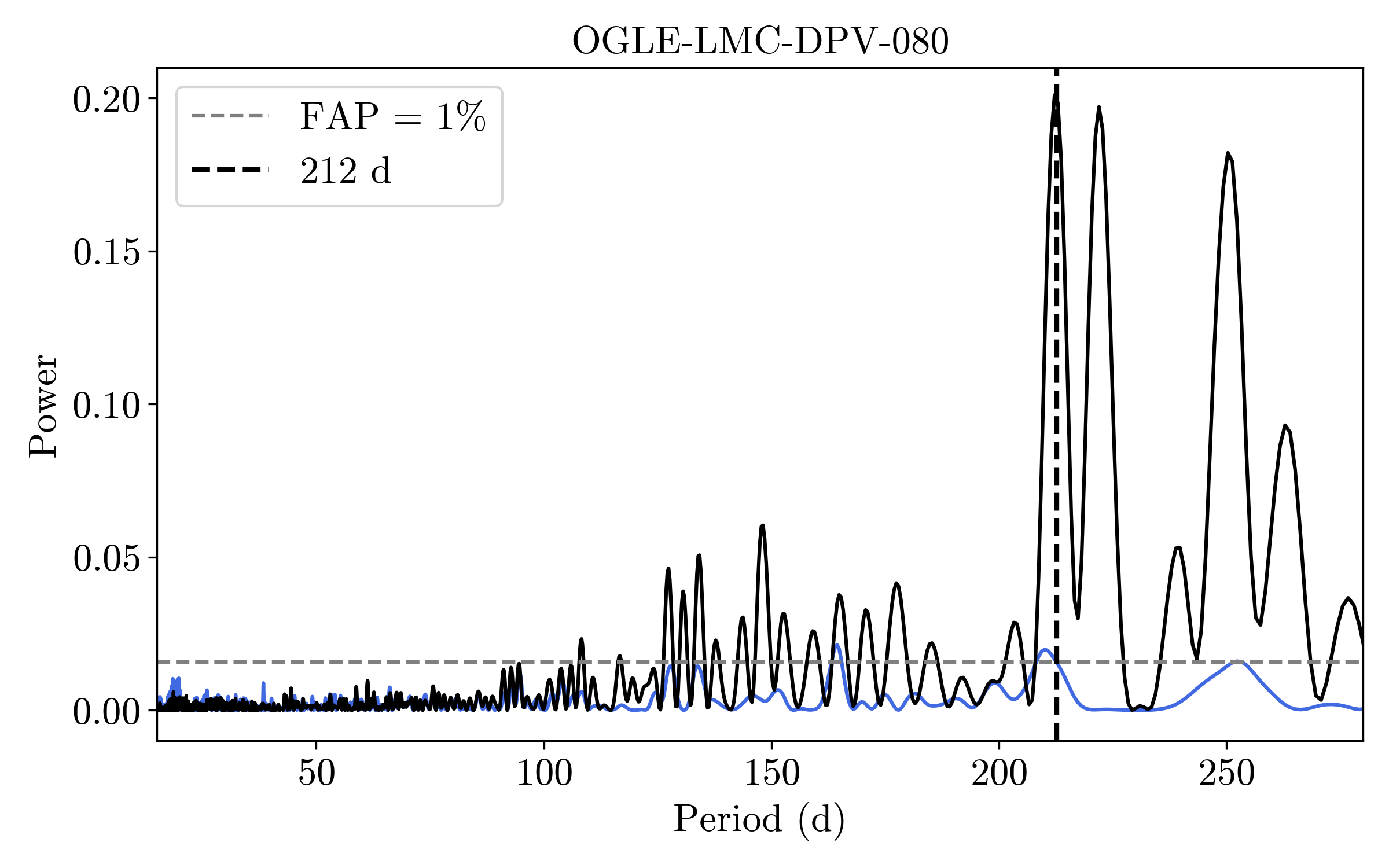}
\end{center}
\caption{GLS periodogram of the residuals of OGLE-LMC-DPV-080 after separating the light curve using a constant (black line) and a variable long period (blue line).}
\label{GLS-080}
\end{figure}

\subsection{Additional frequencies}
\label{Add-freq-sec}

The additional frequencies previously reported in the LMC DPV catalogue \citep{Poleski2010} reappear in our analysis when the light curves of the 18 DPVs with variable long-cycle periods are separated using a constant period. When a variable long period is adopted during the disentangling process, however, these peaks completely disappear from the residuals. This suggests that they are artefacts introduced by assuming a constant long period.

As an example to support this statement, we show in Fig.~\ref{GLS-080} the periodogram of the residuals after we extracted the long cycle and the orbital cycle from the photometry of OGLE-LMC-DPV-080. The black line shows the periodogram of the residuals when assuming a constant long period of 227 days, as reported by \citet{Poleski2010}, while the blue line corresponds to the periodogram after considering a variable long period, as in our study (see Sect.~\ref{variable-long-period-sec}). The peak at 212.77 days, previously reported in the remarks\footnote{\url{https://ogledb.astrouw.edu.pl/\~ogle/CVS/\#dpv}} of the LMC DPV catalogue, is also highlighted. When analysing the periodogram for the second case (blue line), we observe that this peak at $\sim$212 days disappears completely, leaving only a less prominent signal near the 1\% FAP threshold. In \href{https://doi.org/10.5281/zenodo.15881667}{Appendix C} (see \href{https://doi.org/10.5281/zenodo.15881667}{Fig.~C.1}), we present the same analysis for the remaining 17 systems with variable long-cycle periods, demonstrating that the additional frequencies in the residuals also disappear when a variable long period is adopted during the disentangling process.

\section{Discussion}
\label{discussion}
\subsection{Possible origin of the frequency combination}

While the physical origin of the observed frequency combinations remains uncertain, we propose that their appearance may be linked to the disentangling process itself. Although we focused on the changes in the depth of the secondary minimum, we note that the morphology of the orbital light curve varies during the long cycle and beyond the orbital phases we analysed. When these variations occur, a single global fit that is based on a fixed number of Fourier harmonics may be unable to capture them adequately. As a result, structured residuals may remain that manifest as frequency combinations, such as $f_c = f_o + f_l$.

We therefore considered the case of OGLE-BLG-ECL-157529 because the long period and its amplitude decrease over time. Along with this, the changes in the orbital light curve weaken as the long cycle weakens, and a corresponding decrease in the amplitude associated with the combination period $P_c = f_c^{-1}$ is observed. This suggests that orbital photometric changes may play a key role in generating $f_c$ in certain DPVs.

\begin{figure}
\begin{center}
\includegraphics[width=0.99\linewidth]{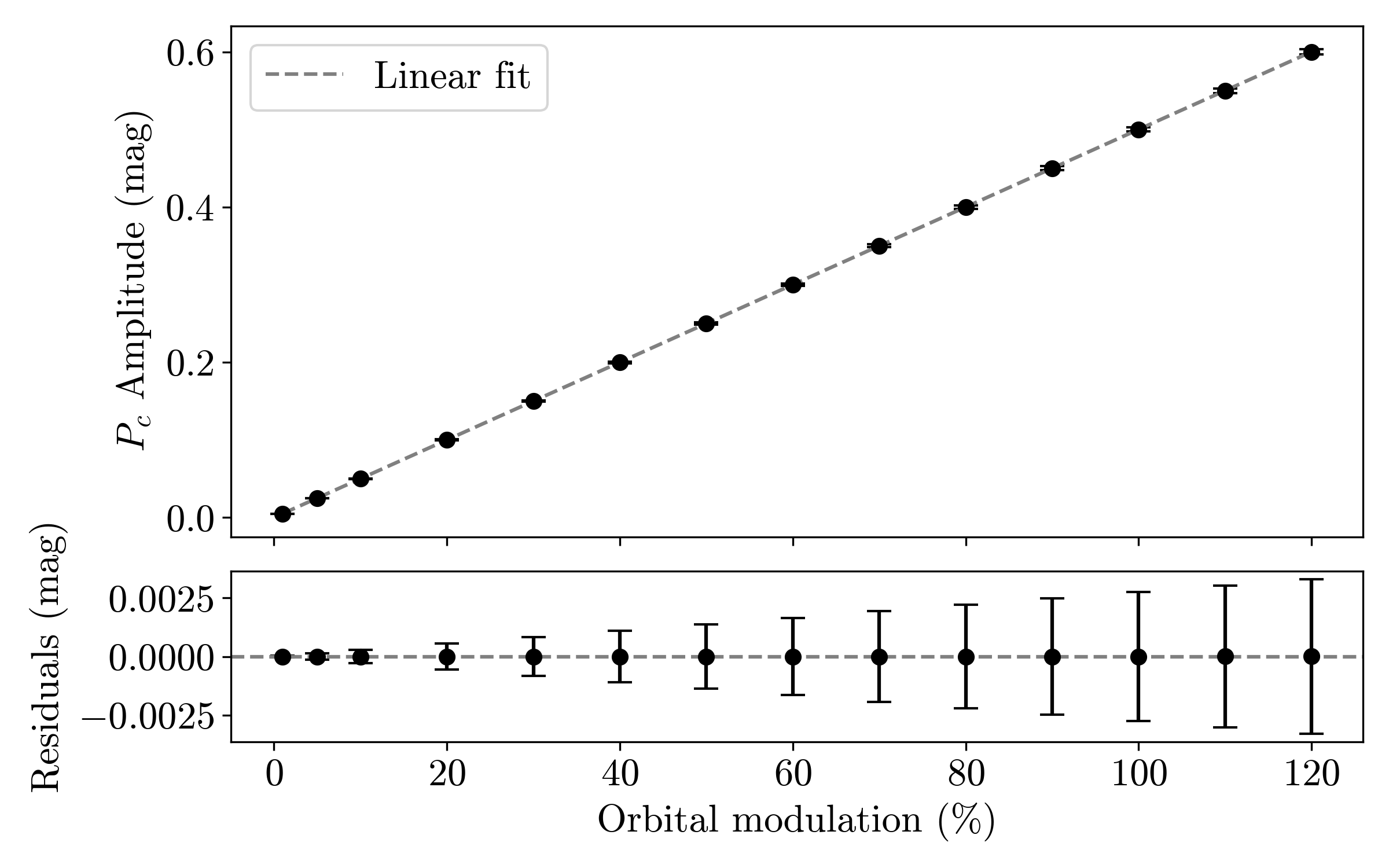}
\end{center}
\caption{Top: Modelled orbital modulation of the orbital light curve vs. the amplitude of the curve with period $P_{c} = f_{c}^{-1}$ found in the residuals. The grey line represents the linear fit. Bottom: Residuals of the linear fit.}
\label{fc-teo}
\end{figure}

To explore this idea, we generated an artificial light curve composed of two sinusoidal components. The first represented the orbital cycle, with a period of one day, and the second represented the long cycle, with a period of 33 days, similar to the observed period ratio in DPVs. The amplitude of the long cycle was set to 20\% of the amplitude of the orbital cycle. To test our hypothesis, we varied the amplitude of the orbital cycle over the long cycle with 1\% to 120\% variations. This mimics the changes observed in the morphology of the orbital light curves of DPVs during the long cycle. After we obtained the final light curve, we proceeded to separate and analyse the residuals and searched for combinations of $f_c$ frequencies that were similar to those found in the DPVs we studied. When the amplitude of the orbital cycle varied cyclically with the long cycle, an $f_c$ frequency always appeared in the residuals. The amplitude of the light curve with period $P_c$ depends directly on the amplitude variation of the orbital cycle throughout the long cycle (see Fig.~\ref{fc-teo}). When the orbital light curve remained stable during the long cycle, no frequency $f_c$ is detected in the residuals. This result supports our hypothesis that the observed frequency combinations $f_c$ are due to cyclic changes in the morphology of the orbital cycle over the long cycle.

\subsection{Implications of the long-cycle variability}

The long cycle in DPVs implies two fundamental components: a clock that regulates its duration, and a physical origin that generates the observed variability. In the context of the magnetic dynamo model proposed by \citet{Schleicher2017}, the clock might be linked to the magnetic cycle of the donor star, and the origin of the cycle might be associated with structural changes in the accretion disc that is modulated by variations in the mass-transfer rate.

Our results show that the long-cycle period is varies in a significant number of DPVs. In several systems, this period varies erratically and alternates between phases of increase and decrease. The most striking aspect, however, is that in most of these cases, the shape and amplitude of the long-cycle light curve remain nearly unchanged (see \href{https://doi.org/10.5281/zenodo.15881667}{Fig.~B.1}, middle and right panels). This behaviour was previously reported for OGLE-LMC-DPV-065 \citep{Mennickent2019} and suggests a decoupling between the internal clock and the physical response underlying the origin of the long cycle. The only known exception so far is OGLE-BLG-ECL-157529, in which the period and amplitude of the long cycle decrease simultaneously \citep{Mennickent2020, Mennickent2021}. This system may represent a stage in which the long cycle vanishes. Its singularity stands out precisely by its contrast with the general behaviour that is observed in the rest of the systems we analysed.

These findings highlight the importance of long-term photometric monitoring to track the evolution of the long cycle on timescales of decades. Systems such as OGLE-BLG-ECL-157529 and OGLE-LMC-DPV-065 are particularly valuable for understanding whether the long cycle stabilises, vanishes, or evolves in more complex ways over time.

\subsection{Origin of the long cycle}

As we demonstrated, the light curves of OGLE-LMC-DPV-097 and OGLE-BLG-ECL-157529 change most strongly throughout the long cycle. The two systems have previously been modelled in detail \citep{Garces2018, Mennickent2021} and therefore provide a solid basis for a direct comparison and for exploring a possible origin of their long-cycle variability.

\begin{figure}
\begin{center}
\includegraphics[width=0.98\linewidth]{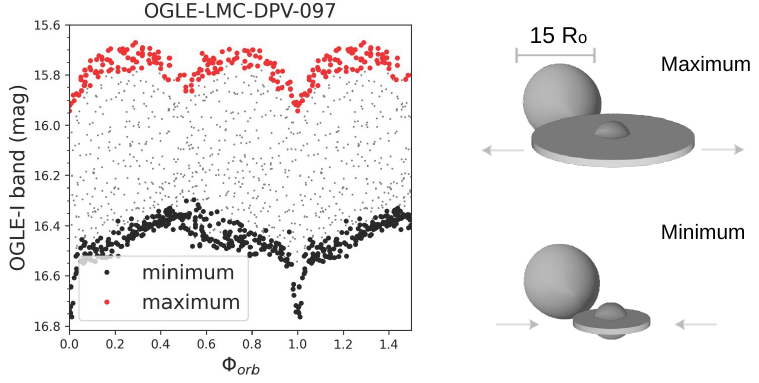}
\includegraphics[width=0.98\linewidth]{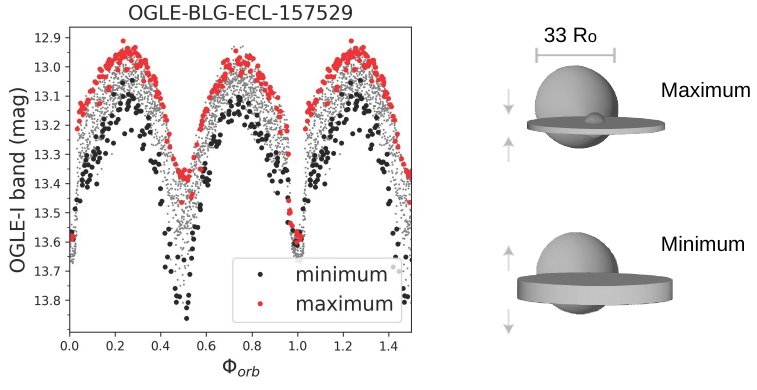}
\end{center}
\caption{Top left: Orbital phase vs. magnitude in the OGLE {\it I} band for OGLE-LMC-DPV-097. The red points show the maximum of the long cycle, and the black points show the minimum. The grey dots represent other phases of the long cycle. Top right: Comparative illustration of the binary system at orbital phase 0.55 in the maximum and minimum phases of the long cycle based on the orbital light-curve models of \citet{Garces2018}. Bottom left: Same as above, but for OGLE-BLG-ECL-157529. Bottom right: Behaviour of the accretion disc in OGLE-BLG-ECL-157529 according to the best-fitting orbital light-curve models LC-4 (long-cycle minimum) and LC-7 (maximum) from \citet[see Table~A.1 of that study]{Mennickent2021}.}
\label{cam-lc}
\end{figure}

In OGLE-LMC-DPV-097, a comparison of the disc structure in opposite phases of the long cycle revealed that the disc is approximately 50\% smaller but 2.3 times hotter during the long-cycle minimum \citep{Garces2018}. This behaviour strongly affects the orbital light curve, in particular, around the secondary minimum (orbital phase 0.5). During the long-cycle minimum, the smaller disc and the lower orbital inclination of the system prevent the donor from being eclipsed. Additionally, the bright spot in the disc enhances its luminosity, which leads to a maximum brightness near phase 0.5 (see the upper left panel of Fig.~\ref{cam-lc}). In contrast, during the long-cycle maximum, the disc expands radially and partially eclipses the donor, resulting in a deeper secondary minimum. In this system, the disc together with the orbital inclination ($i = 74^{\circ}$) therefore drives the long-cycle modulation and the changes we observed in the orbital light curves. Moreover, a similar behaviour of the disc, if more subtle, might explain the deeper secondary minimum observed during long-cycle maximum in 38\% of the LMC DPVs we analysed. This group includes OGLE-LMC-DPV-065, for which a recent study suggested that structural changes in the accretion disc can reproduce the orbital and long photometric cycles. It also demonstrated that the disc becomes more extended radially during the long-cycle maximum \citep{MennickentDjurasevic2025}.

The most significant changes in OGLE-BLG-ECL-157529 occur in the central thickness of the accretion disc, which becomes 1.4 times thicker during the long-cycle minimum than at the maximum, while the disc temperature and radius remain nearly constant \citep{Mennickent2021}. As a result, a larger surface area of the donor star is occulted during the long-cycle minimum, which leads to a deeper secondary minimum. In contrast, the thinner disc and a smaller portion of the donor partially occult the gainer during the long-cycle maximum, resulting in a brighter secondary minimum (see lower left panel of Fig.~\ref{cam-lc}). For this system, the long cycles are therefore explained by occultation of the gainer as a result of the changes in the vertical disc extension, as proposed by \citet{Mennickent2021}. A similar but less pronounced behaviour might account for the variability associated with a deeper secondary minimum during the long-cycle minimum that is observed in approximately 12\% of the LMC DPVs we analysed. Notably, OGLE-LMC-DPV-074 is one such case, and recent work showed that the changes in its accretion disc throughout the long cycle follow the same pattern as was observed in OGLE-BLG-ECL-157529 \citep{Mennickent2025}.

In principle, edge-on discs are expected to primarily affect the long cycle through changes in the thickness that occult the gainer, while discs observed at intermediate inclinations contribute to the long cycle through changes in their radial size and projected area along the line of sight. In this scenario, we would be observing the same underlying phenomenon (discs that vary in the vertical and radial extension), but seen at different inclinations. This offers a general explanation for all observed cases. In most LMC DPVs, the changes in the morphology of the orbital light curves are less drastic. It is therefore crucial to understanding the long cycle in DPVs and to testing the above scenario to understand what occurs in the accretion discs of OGLE-LMC-DPV-097 and OGLE-BLG-ECL-157529. The determination of the disc configuration and system inclination proves to be decisive for this goal and also for explaining the similar long-cycle amplitude distribution in ellipsoidal and eclipsing DPVs while no drastic changes are observed in their orbital light curves during the long cycle (see Fig.~\ref{delta-mag}, Sect.~\ref{cambios-curvas}).

Along with the results presented here, recent studies supported the idea that the photometric long cycle in DPVs might originate either directly in intrinsic changes in the accretion disc, as seen in OGLE-LMC-DPV-097 and OGLE-LMC-DPV-065, or in the partial occultation of the gainer as a consequence of disc changes, as proposed for OGLE-BLG-ECL-157529 and OGLE-LMC-DPV-074. In both scenarios, the mass-transfer rate might play a key role in modulating the disc structure and vertical extension. It remains to be investigated whether a common underlying mechanism causes the different reported disc behaviour, however.

\section{Conclusion}
\label{conclusion}

The photometric analysis of the orbital light curves of the LMC DPVs we studied here led to the conclusions listed below.

First, when a $3\sigma$ significance criterion was applied, the behaviour of 38\% of the LMC DPVs was similar to that of OGLE-LMC-DPV-097, which is characterised by a deeper secondary minimum during the long-cycle maximum. In contrast, 12\% of the LMC DPVs showed the opposite behaviour and were similar to OGLE-BLG-ECL-157529, which is defined by a deeper secondary minimum during the long-cycle minimum. Both OGLE-LMC-DPV-097 and OGLE-BLG-ECL-157529 are extreme cases with evident changes in the morphology of the orbital light curves in opposite phases of the long cycle. In addition, 50\% of the DPVs changed only slightly in the depth of the secondary minimum, and 93\% of these are classified as ellipsoidal DPVs.

Second, we identified 18 systems with a variable long period, 10 of which are new cases. The long period in 2 DPVs increased over time, decreased for 5 objects, and was mixed for 11 objects. Our analysis of the rms confirmed that the long-cycle light curves maintain their morphology despite changes in the period in all 18 cases we studied. We measured the rms of the long-cycle light curves before and after we accounted for a variable period in the disentangling process and observed a reduced dispersion of between 15\% and 62\% relative to the initial rms value. Along with the time–period and GLS analysis, this confirmed that the long period in these systems is variable.

Third, we increased the number of DPVs with frequency combinations in their residuals to 73. We estimated the amplitude of the light curve with a combination period $P_c = f_{c}^{-1}$ in the residuals and found a positive correlation between the intensity of the depth change in the secondary minimum of the orbital light curve and its amplitude. This correlation appears to be stronger for eclipsing DPVs (when OGLE-LMC-DPV-097 is included) than for ellipsoidal DPVs. We also found that the additional frequencies that were reported for some LMC DPVs disappear when a variable long period was considered in the disentangling process. It is therefore reasonable to conclude that these frequencies are artefacts produced by variability associated with the long-cycle duration.

Finally, we showed that the extreme cases of OGLE-BLG-ECL-157529 and OGLE-LMC-DPV-097 represent systems that are observed at medium, high, and very high inclinations, where the long cycle can be explained as the result of gainer occultation due to a the variable vertical thickness and variable radial extension of a disc, respectively. It remains to be tested through future light-curve modelling, whether systems with similar light curves can be explained by the same phenomenon. In any case, the changes observed in the accretion discs of DPVs require an explanation. One possible cause is a variable mass-transfer rate that regulates the disc structure through the effect on the disc and the subsequent reorganisation of material around the gainer. The cause of the variable stream remains unknown, but the previously proposed hypothesis of a donor magnetic dynamo that cyclically regulates the mass transfer is a feasible scenario that merits confirmation.

\section*{Data availability}
An external appendix containing complementary figures for Sects.~\ref{variable-long-period-sec} and \ref{Add-freq-sec} is available on Zenodo at \url{https://doi.org/10.5281/zenodo.15881667}.

\begin{acknowledgements}
We acknowledge the comments and suggestions of the anonymous referee who reviewed the first version of this article. J.G. acknowledges support by ANID grant Nro. 21202285. R.M., J.G. and D.R.G.S. acknowledge support by the ANID BASAL project Centro de Astrof{\'{i}}sica y Tecnolog{\'{i}}as Afines ACE210002 (CATA). J.P. acknowledges the Ministry of Science, Technological Development, and Innovations of the Republic of Serbia for their support. D.B. acknowledges support by Fondecyt grant Nro. 11230261. L.C. acknowledges economic support from ANID-Subdireccion de capital humano/doctorado nacional/2022-21220607. C.C.C. thanks for Fondo IPIG-ACA-CaCo financiado por el instrumento para la igualdad de género de la Universidad de Talca, Proyecto InES-Género INGE210025, Agencia nacional de Investigación y Desarrollo (ANID). D.R.G.S. thanks for funding via the Alexander von Humboldt - Foundation, Bonn, Germany. I.S.: This work has been co-funded by the National Science Centre, Poland, grant No. 2022/45/B/ST9/00243.
\end{acknowledgements}

\bibliography{biblio}
\bibliographystyle{aa}

\clearpage
\onecolumn
\begin{appendix}
\label{Appendix}
\section{Additional figures and summary tables.}
\FloatBarrier
\begin{figure*}[h!]
\begin{center}
\begin{tabular}{cl}
\includegraphics[width=0.48\linewidth]{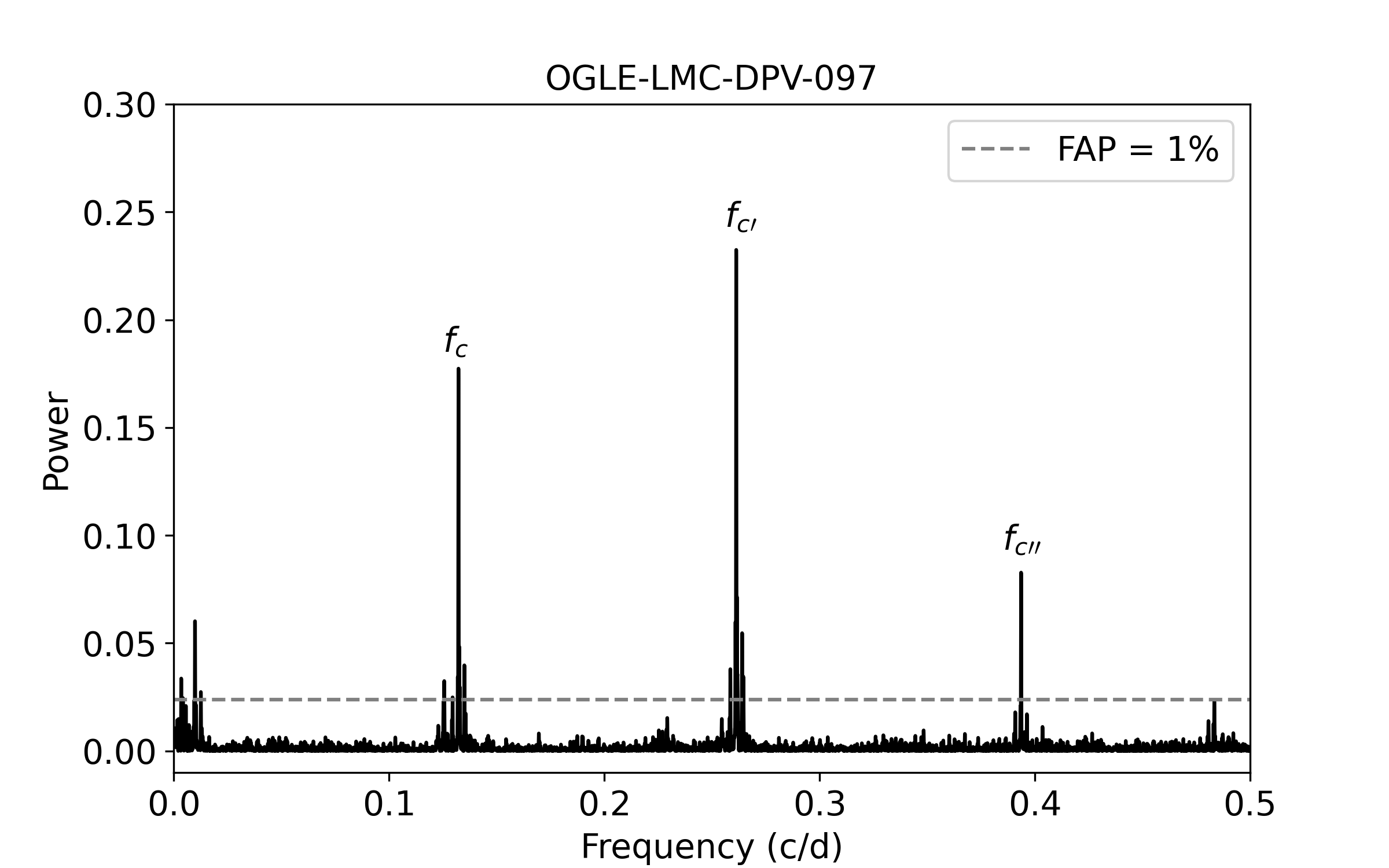}&
\includegraphics[width=0.48\linewidth]{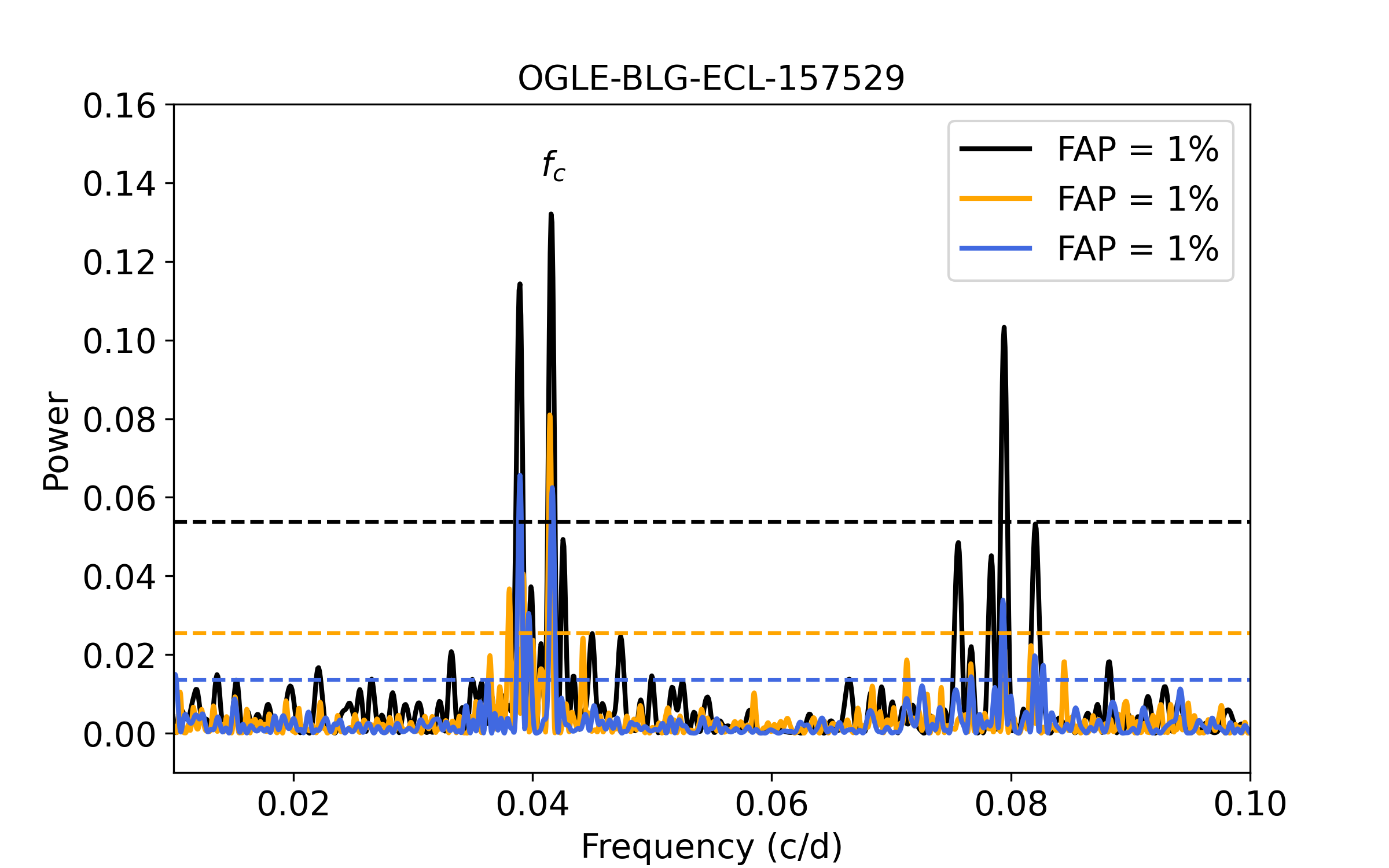}
\end{tabular}
\end{center}
\caption{Left: GLS periodogram of the residual data of OGLE-LMC-DPV-097. Different frequency combinations are shown: $f_{c} = f_{o} + f_{l}$, $f_{c'} = 2f_o + f_l$, and $f_{c''} = 3f_o + 2f_l$. Right: GLS periodogram of the OGLE-BLG-ECL-157529 residuals for different sets of OGLE {\it I} band photometric data. The first set ranges from HJD 500 to 4000 (black line), the second from 4000 to 5050 (orange line), and the last from 5060 to 7500 (blue line). For more details, see Section~\ref{freq-in-res}.}
\label{Per-cambios}
\end{figure*}

\FloatBarrier

\onecolumn
\begin{longtable}{lcclcc}
\caption{Summary of the analysis of the residuals using the GLS periodogram.}\\
\label{tab-fre-com}\\
\hline
\hline
OGLE-ID               &       $f_{c}$            &  Amplitude  &  OGLE-ID               &       $f_{c}$            &  Amplitude     \\
\hline
\endfirsthead
\caption{continued.}\\
\hline
\hline
OGLE-ID               &       $f_{c}$            &  Amplitude  &  OGLE-ID               &       $f_{c}$            &  Amplitude     \\
\hline
\endhead
\hline
\endfoot
OGLE-LMC-DPV-002                & $ f_{o} + f_{l} $  &  0.024 $\pm$ 0.003 &  OGLE-LMC-DPV-062   & $2f_{o} + f_{l}$  &  0.021 $\pm$ 0.001 \\
OGLE-LMC-DPV-004                & $ f_{o} + f_{l} $  &  0.011 $\pm$ 0.002 &  OGLE-LMC-DPV-063   & $f_{o} - f_{l}$   &  0.018 $\pm$ 0.003 \\
OGLE-LMC-DPV-006            	& $ 2f_{o} + f_{l}$  &  0.022 $\pm$ 0.002 &  OGLE-LMC-DPV-064$^{*, \dagger}$ &      $   f_{o} + f_{l}  $   &  0.012 $\pm$    0.001   \\
OGLE-LMC-DPV-007            	&      $   2f_{o} - f_{l}       $  &  0.019 $\pm$    0.002    &  OGLE-LMC-DPV-066$^{*}$  &      $   f_{o} + f_{l}  $ &  0.019 $\pm$    0.002  \\
OGLE-LMC-DPV-011            	&      $   f_{o} + f_{l}        $  &  0.009 $\pm$    0.002   &  OGLE-LMC-DPV-067$^{*, \dagger}$ &      $   f_{o} + f_{l}  $  &  0.018 $\pm$    0.001  \\
OGLE-LMC-DPV-013            	&      $   f_{o} + f_{l}        $  &  0.011 $\pm$    0.002    &  OGLE-LMC-DPV-069$^{*}$    &      $   f_{o} + f_{l}  $  &  0.021 $\pm$    0.002  \\
OGLE-LMC-DPV-014            	&      $   f_{o} + f_{l}        $  &  0.012 $\pm$    0.002     &  OGLE-LMC-DPV-070  &      $   f_{o} + f_{l}    $  &  0.017 $\pm$    0.002        \\
OGLE-LMC-DPV-015$^{*}$      	&      $   f_{o} + f_{l}        $  &  0.019 $\pm$    0.001     &  OGLE-LMC-DPV-073   &      $   3f_{o} + f_{l} $  &  0.032 $\pm$    0.003        \\
OGLE-LMC-DPV-016            	&      $   f_{o} + f_{l}        $  &  0.006 $\pm$    0.001     &  OGLE-LMC-DPV-074   &      $   f_{o} + f_{l}        $  &  0.025 $\pm$    0.002 \\
OGLE-LMC-DPV-018            	&      $   2f_{o}               $  &  0.031 $\pm$    0.004     &  OGLE-LMC-DPV-078$^{*}$   &  $   f_{o} + f_{l}  $  &  0.014 $\pm$    0.001  \\
OGLE-LMC-DPV-019$^{*}$      	&      $   f_{o} + f_{l}  $        &  0.014 $\pm$    0.001   &	    OGLE-LMC-DPV-079  &      $   f_{o} + f_{l}        $  &  0.013 $\pm$    0.002        \\
OGLE-LMC-DPV-020$^{*}$      	&      $   f_{o} + f_{l}  $        &  0.011 $\pm$    0.001   &    OGLE-LMC-DPV-081   &  $   f_{o} + f_{l}  $  &  0.018 $\pm$     0.002 \\
OGLE-LMC-DPV-021$^{*}$      	&      $   f_{o} + f_{l}  $        &  0.017 $\pm$    0.003   &	    OGLE-LMC-DPV-083$^{*}$ &      $   f_{o} + f_{l}  $   &  0.009 $\pm$  0.001  \\
OGLE-LMC-DPV-022            	&      $   3f_{o} - f_{l}       $  &  0.020 $\pm$    0.002    &  OGLE-LMC-DPV-084$^{*, \dagger}$ &   $   f_{o} + f_{l}  $ &  0.021 $\pm$    0.001  \\
OGLE-LMC-DPV-024            	&      $   2f_{o} + f_{l}       $  &  0.013 $\pm$    0.002     &  OGLE-LMC-DPV-089$^{*, \dagger}$ &   $   f_{o} + f_{l}  $  &  0.014 $\pm$    0.001  \\
OGLE-LMC-DPV-025$^{\dagger}$    &      $   f_{o} + f_{l}        $  &  0.021 $\pm$    0.001      &  OGLE-LMC-DPV-090   &  $   f_{o} + f_{l}  $  &  0.022 $\pm$     0.002 \\
OGLE-LMC-DPV-026                &      $   2f_{o} + f_{l}       $  &  0.019 $\pm$    0.001      &  OGLE-LMC-DPV-097$^{*}$   & $   f_{o} + f_{l}  $   &  0.045 $\pm$    0.003  \\
OGLE-LMC-DPV-027$^{*}$      	&      $   f_{o} + f_{l}  $        &  0.013 $\pm$    0.001   &	    OGLE-LMC-DPV-098$^{*}$    &      $   f_{o} + f_{l}  $    &  0.015 $\pm$    0.002  \\
OGLE-LMC-DPV-029$^{*}$      	&      $   f_{o} + f_{l}  $        &  0.014 $\pm$    0.001   &	    OGLE-LMC-DPV-101$^{*}$      	&  $   f_{o} + f_{l}  $    &  0.011 $\pm$    0.002  \\
OGLE-LMC-DPV-031            	&      $   f_{o} + f_{l}        $  &  0.008 $\pm$    0.002    &  OGLE-LMC-DPV-102$^{*}$  	&      $   f_{o} + f_{l}  $  &  0.011 $\pm$    0.001  \\
OGLE-LMC-DPV-032            	&      $   2f_{o} + f_{l}       $  &  0.018 $\pm$    0.001     &      OGLE-LMC-DPV-104  &      $   f_{o} + f_{l}        $  &  0.011 $\pm$    0.002        \\
OGLE-LMC-DPV-034$^{*}$      	&      $   f_{o} - f_{l}        $  &  0.025 $\pm$    0.002    &    OGLE-LMC-DPV-105$^{*}$  &  $   f_{o} + f_{l} $ & 0.025 $\pm$  0.001  \\
OGLE-LMC-DPV-033   &  $   f_{o} + f_{l}  $        &  0.009 $\pm$     0.002                    &    OGLE-LMC-DPV-107$^{\dagger}$    &   $   f_{o} + f_{l}   $  &  0.011 $\pm$    0.002   \\
OGLE-LMC-DPV-035$^{*, \dagger}$ &      $   f_{o} + f_{l}  $        &  0.022 $\pm$    0.002   &    OGLE-LMC-DPV-108   &  $   f_{o} + f_{l}  $   &  0.014 $\pm$     0.001  \\
OGLE-LMC-DPV-037   &  $   f_{o} - f_{l}  $        &  0.009 $\pm$     0.002  &  OGLE-LMC-DPV-109     &      $   2f_{o} + f_{l}       $  &  0.016 $\pm$    0.002       \\
OGLE-LMC-DPV-038$^{*}$          &      $   f_{o} + f_{l}  $        &  0.014 $\pm$    0.002   &            OGLE-LMC-DPV-110   &  $   f_{o} + f_{l}  $        &  0.012 $\pm$     0.002 \\
OGLE-LMC-DPV-039            	&      $   f_{o} + f_{l}        $  &  0.007 $\pm$    0.002  &  OGLE-LMC-DPV-111  	&      $   f_{o} + f_{l}   $  &  0.013 $\pm$    0.001       \\
OGLE-LMC-DPV-043            	&      $   f_{o} + f_{l}        $  &  0.007 $\pm$    0.002     &  OGLE-LMC-DPV-113    	&      $   f_{o} + f_{l}   $  &  0.013 $\pm$    0.002        \\
OGLE-LMC-DPV-047   &  $   f_{o} + f_{l}  $        &  0.006 $\pm$     0.001 &   OGLE-LMC-DPV-115$^{*}$   	&      $f_{o} + f_{l}  $        &  0.018 $\pm$    0.001  \\
OGLE-LMC-DPV-048   &  $   f_{o} - f_{l}  $        &  0.012 $\pm$     0.001  &  OGLE-LMC-DPV-116$^{*}$ 	&    $   f_{o} + f_{l}  $   &  0.010 $\pm$    0.001  \\
OGLE-LMC-DPV-050   &  $   f_{o} + f_{l}  $        &  0.008 $\pm$     0.002  &  OGLE-LMC-DPV-117      &    $   f_{o} + f_{l} $  &  0.009 $\pm$    0.002    \\
OGLE-LMC-DPV-051            	&      $   2f_{o} + f_{l}       $  &  0.019 $\pm$    0.002   &  OGLE-LMC-DPV-118$^{*}$	& $   f_{o} + f_{l}  $ &  0.013 $\pm$    0.001  \\
OGLE-LMC-DPV-052   &  $   2f_{o} + f_{l}  $       &  0.018 $\pm$     0.004  &		    OGLE-LMC-DPV-119$^{*}$   &  $   f_{o} + f_{l}  $   &  0.012 $\pm$    0.002  \\
OGLE-LMC-DPV-053            	&      $   3f_{o} + f_{l}       $  &  0.026 $\pm$    0.002      & OGLE-LMC-DPV-121$^{*}$  &  $   f_{o} + f_{l}  $        &  0.017 $\pm$    0.003  \\
OGLE-LMC-DPV-056   &  $   f_{o} + f_{l}  $        &  0.008 $\pm$     0.003  &  OGLE-LMC-DPV-123$^{*}$ &   $   f_{o} + f_{l}  $        &  0.017 $\pm$    0.004  \\
OGLE-LMC-DPV-060            	&      $   2f_{o} + f_{l}       $  &  0.007 $\pm$    0.001     & OGLE-LMC-DPV-134$^{*}$ &   $   f_{o} - f_{l}  $  &  0.013 $\pm$    0.002        \\
OGLE-LMC-DPV-061            	&      $   2f_{o} - 2f          $  &  0.011 $\pm$    0.001        \\
\end{longtable}
\tablefoot{We list the OGLE ID and the amplitude of the light curve associated with the frequency combination $f_c = f_o \pm f_l$, where $f_o$ and $f_l$ are the orbital and long-cycle frequencies, respectively. OGLE IDs marked with $*$ and $\dagger$ correspond to frequency combinations previously reported by \citet{Poleski2010} and \citet{Buchler2009}, respectively.}

\onecolumn
\begin{longtable}{lrclrc}
\caption{OGLE ID, $\Delta A_{II}$, and long-cycle amplitude $A_{LC}$ for each system, grouped by type and significance level.}\\
\label{tab-res-analisis-1}\\
\hline
\hline
\multicolumn{6}{c}{Eclipsing DPVs}\\
\hline
\multicolumn{6}{c}{$|\Delta A_{II}| > 3\sigma$}\\
\hline
OGLE-ID               &   $\Delta A_{II}$           &  $A_{LC}$  &  OGLE-ID               &       $\Delta A_{II}$           &  $A_{LC}$  \\
\hline
\endfirsthead
\caption{continued.}\\
\hline
\hline
\multicolumn{6}{c}{Ellipsoidals DPVs} \\
\hline
\multicolumn{6}{c}{ $|\Delta A_{II}| \leq 3\sigma$} \\
\hline
OGLE-ID               &   $\Delta A_{II}$           &  $A_{LC}$  &  OGLE-ID               &       $\Delta A_{II}$           &  $A_{LC}$  \\
\hline
\endhead
\hline
\endfoot
OGLE-LMC-DPV-005 & $0.039 \pm 0.010$ & $0.118 \pm 0.012$ &  OGLE-LMC-DPV-063 & $0.061 \pm 0.001$ & $0.115 \pm 0.004$ \\
OGLE-LMC-DPV-006 & $0.058 \pm 0.007$ & $0.159 \pm 0.007$ &  OGLE-LMC-DPV-065 & $0.034 \pm 0.009$ & $0.185 \pm 0.015$ \\
OGLE-LMC-DPV-014 & $0.058 \pm 0.007$ & $0.191 \pm 0.006$ &  OGLE-LMC-DPV-073 & $-0.060 \pm 0.010$ & $0.140 \pm 0.004$ \\
OGLE-LMC-DPV-018 & $-0.099 \pm 0.001$ & $0.094 \pm 0.005$ & OGLE-LMC-DPV-074 & $-0.062 \pm 0.006$ & $0.089 \pm 0.017$ \\
OGLE-LMC-DPV-021 & $0.071 \pm 0.001$ & $0.266 \pm 0.009$ &  OGLE-LMC-DPV-097 & $0.193 \pm 0.011$ & $0.756 \pm 0.020$ \\
OGLE-LMC-DPV-024 & $0.054 \pm 0.001$ & $0.074 \pm 0.018$ &  OGLE-LMC-DPV-098 & $0.037 \pm 0.001$ & $0.183 \pm 0.010$ \\
OGLE-LMC-DPV-026 & $0.048 \pm 0.006$ & $0.146 \pm 0.006$ &  OGLE-LMC-DPV-101 & $0.040 \pm 0.009$ & $0.106 \pm 0.018$ \\
OGLE-LMC-DPV-030 & $0.026 \pm 0.001$ & $0.096 \pm 0.012$ &  OGLE-LMC-DPV-108 & $-0.023 \pm 0.001$ & $0.177 \pm 0.004$ \\
OGLE-LMC-DPV-032 & $0.038 \pm 0.009$ & $0.230 \pm 0.017$ &  OGLE-LMC-DPV-109 & $0.068 \pm 0.001$ & $0.122 \pm 0.011$ \\
OGLE-LMC-DPV-034 & $-0.026 \pm 0.001$ & $0.108 \pm 0.019$ & OGLE-LMC-DPV-112 & $0.029 \pm 0.001$ & $0.135 \pm 0.018$ \\
OGLE-LMC-DPV-049 & $0.070 \pm 0.007$ & $0.390 \pm 0.018$ &  OGLE-LMC-DPV-121 & $0.089 \pm 0.008$ & $0.267 \pm 0.016$ \\
OGLE-LMC-DPV-051 & $0.032 \pm 0.001$ & $0.273 \pm 0.002$ &  OGLE-LMC-DPV-124 & $0.034 \pm 0.001$ & $0.067 \pm 0.014$ \\
OGLE-LMC-DPV-053 & $-0.069 \pm 0.001$ & $0.063 \pm 0.012$ & OGLE-LMC-DPV-131 & $0.030 \pm 0.001$ & $0.022 \pm 0.013$ \\
OGLE-LMC-DPV-056 & $0.050 \pm 0.005$ & $0.222 \pm 0.007$ &  OGLE-LMC-DPV-134 & $-0.014 \pm 0.001$ & $0.066 \pm 0.012$ \\
OGLE-LMC-DPV-058 & $0.041 \pm 0.005$ & $0.249 \pm 0.004$ &  OGLE-LMC-DPV-137 & $0.017 \pm 0.001$ & $0.028 \pm 0.009$ \\
OGLE-LMC-DPV-062 & $0.046 \pm 0.008$ & $0.190 \pm 0.006$ &        &             &            \\
\hline
\multicolumn{6}{c}{$|\Delta A_{II}| \leq 3\sigma$}\\
\hline
OGLE-LMC-DPV-012 & $0.019 \pm 0.008$ & $0.074 \pm 0.007$ &  OGLE-LMC-DPV-061 & $0.003 \pm 0.007$ & $0.049 \pm 0.014$ \\
OGLE-LMC-DPV-040 & $0.004 \pm 0.009$ & $0.060 \pm 0.006$ &  OGLE-LMC-DPV-129 & $0.013 \pm 0.007$ & $0.031 \pm 0.010$ \\
OGLE-LMC-DPV-060 & $0.022 \pm 0.008$ & $0.058 \pm 0.005$ &  & & \\
\hline
\multicolumn{6}{c}{Ellipsoidals DPVs} \\
\hline
\multicolumn{6}{c}{ $|\Delta A_{II}| > 3\sigma$} \\
\hline
OGLE-ID               &   $\Delta A_{II}$            &  $A_{LC}$  &  OGLE-ID               &       $\Delta A_{II}$           &  $A_{LC}$  \\
\hline
OGLE-LMC-DPV-001 & $0.041 \pm 0.012$ & $0.541 \pm 0.018$ &  OGLE-LMC-DPV-075 & $0.068 \pm 0.002$ & $0.072 \pm 0.006$ \\
OGLE-LMC-DPV-007 & $0.030 \pm 0.007$ & $0.167 \pm 0.009$ &  OGLE-LMC-DPV-079 & $0.029 \pm 0.004$ & $0.325 \pm 0.009$ \\
OGLE-LMC-DPV-015 & $0.022 \pm 0.005$ & $0.136 \pm 0.003$ &  OGLE-LMC-DPV-081 & $0.051 \pm 0.008$ & $0.436 \pm 0.013$ \\
OGLE-LMC-DPV-020 & $0.021 \pm 0.007$ & $0.210 \pm 0.007$ &  OGLE-LMC-DPV-083 & $0.018 \pm 0.004$ & $0.169 \pm 0.002$ \\
OGLE-LMC-DPV-023 & $-0.008 \pm 0.001$ & $0.057 \pm 0.007$ & OGLE-LMC-DPV-084 & $-0.019 \pm 0.005$ & $0.137 \pm 0.005$ \\
OGLE-LMC-DPV-025 & $0.032 \pm 0.006$ & $0.369 \pm 0.008$ &  OGLE-LMC-DPV-090 & $0.040 \pm 0.009$ & $0.330 \pm 0.010$ \\
OGLE-LMC-DPV-027 & $0.006 \pm 0.001$ & $0.225 \pm 0.004$ &  OGLE-LMC-DPV-096 & $-0.015 \pm 0.001$ & $0.023 \pm 0.004$ \\
OGLE-LMC-DPV-029 & $0.018 \pm 0.003$ & $0.089 \pm 0.002$ &  OGLE-LMC-DPV-099 & $-0.006 \pm 0.001$ & $0.078 \pm 0.003$ \\
OGLE-LMC-DPV-035 & $0.034 \pm 0.005$ & $0.379 \pm 0.005$ &  OGLE-LMC-DPV-103 & $-0.026 \pm 0.006$ & $0.100 \pm 0.016$ \\
OGLE-LMC-DPV-036 & $-0.031 \pm 0.005$ & $0.157 \pm 0.007$ & OGLE-LMC-DPV-105 & $0.032 \pm 0.001$ & $0.175 \pm 0.007$ \\
OGLE-LMC-DPV-037 & $0.018 \pm 0.001$ & $0.045 \pm 0.008$ &  OGLE-LMC-DPV-107 & $0.047 \pm 0.004$ & $0.190 \pm 0.009$ \\
OGLE-LMC-DPV-038 & $0.016 \pm 0.004$ & $0.205 \pm 0.004$ &  OGLE-LMC-DPV-111 & $0.015 \pm 0.004$ & $0.163 \pm 0.004$ \\
OGLE-LMC-DPV-045 & $0.010 \pm 0.003$ & $0.118 \pm 0.007$ &  OGLE-LMC-DPV-114 & $-0.007 \pm 0.001$ & $0.253 \pm 0.002$ \\
OGLE-LMC-DPV-050 & $0.010 \pm 0.002$ & $0.047 \pm 0.003$ &  OGLE-LMC-DPV-117 & $0.005 \pm 0.001$ & $0.127 \pm 0.006$ \\
OGLE-LMC-DPV-064 & $0.018 \pm 0.005$ & $0.099 \pm 0.005$ &  OGLE-LMC-DPV-118 & $0.019 \pm 0.005$ & $0.047 \pm 0.003$ \\
OGLE-LMC-DPV-066 & $0.038 \pm 0.008$ & $0.430 \pm 0.006$ &  OGLE-LMC-DPV-120 & $0.048 \pm 0.001$ & $0.141 \pm 0.003$ \\
OGLE-LMC-DPV-070 & $0.030 \pm 0.001$ & $0.105 \pm 0.003$ &  OGLE-LMC-DPV-123 & $-0.015 \pm 0.001$ & $0.275 \pm 0.015$ \\
OGLE-LMC-DPV-071 & $-0.058 \pm 0.005$ & $0.051 \pm 0.002$ & OGLE-LMC-DPV-135 & $0.011 \pm 0.001$ & $0.040 \pm 0.003$ \\
\hline
\multicolumn{6}{c}{ $|\Delta A_{II}| \leq 3\sigma$} \\
\hline
OGLE-LMC-DPV-002 & $0.019 \pm 0.013$ & $0.195 \pm 0.008$ &  OGLE-LMC-DPV-076 & $-0.011 \pm 0.008$ & $0.059 \pm 0.004$ \\
OGLE-LMC-DPV-003 & $0.020 \pm 0.013$ & $0.060 \pm 0.010$ &  OGLE-LMC-DPV-077 & $0.009 \pm 0.005$ & $0.092 \pm 0.004$ \\
OGLE-LMC-DPV-004 & $0.016 \pm 0.013$ & $0.185 \pm 0.005$ &  OGLE-LMC-DPV-078 & $0.015 \pm 0.006$ & $0.145 \pm 0.004$ \\
OGLE-LMC-DPV-008 & $0.012 \pm 0.005$ & $0.035 \pm 0.002$ &  OGLE-LMC-DPV-080 & $0.004 \pm 0.008$ & $0.098 \pm 0.003$ \\
OGLE-LMC-DPV-009 & $-0.004 \pm 0.007$ & $0.093 \pm 0.006$ & OGLE-LMC-DPV-082 & $0.012 \pm 0.005$ & $0.061 \pm 0.006$ \\
OGLE-LMC-DPV-010 & $0.000 \pm 0.006$ & $0.077 \pm 0.008$ &  OGLE-LMC-DPV-085 & $0.012 \pm 0.004$ & $0.028 \pm 0.008$ \\
OGLE-LMC-DPV-011 & $-0.001 \pm 0.012$ & $0.099 \pm 0.008$ & OGLE-LMC-DPV-086 & $-0.002 \pm 0.004$ & $0.085 \pm 0.005$ \\
OGLE-LMC-DPV-013 & $0.015 \pm 0.009$ & $0.116 \pm 0.008$ &  OGLE-LMC-DPV-087 & $0.011 \pm 0.010$ & $0.075 \pm 0.008$ \\
OGLE-LMC-DPV-016 & $-0.004 \pm 0.007$ & $0.117 \pm 0.005$ & OGLE-LMC-DPV-088 & $-0.005 \pm 0.008$ & $0.067 \pm 0.003$ \\
OGLE-LMC-DPV-017 & $-0.002 \pm 0.008$ & $0.103 \pm 0.012$ & OGLE-LMC-DPV-089 & $0.017 \pm 0.006$ & $0.212 \pm 0.004$ \\
OGLE-LMC-DPV-019 & $0.005 \pm 0.004$ & $0.248 \pm 0.004$ &  OGLE-LMC-DPV-091 & $0.016 \pm 0.007$ & $0.056 \pm 0.017$ \\
OGLE-LMC-DPV-022 & $-0.015 \pm 0.009$ & $0.126 \pm 0.004$ & OGLE-LMC-DPV-092 & $0.011 \pm 0.005$ & $0.075 \pm 0.006$ \\
OGLE-LMC-DPV-028 & $0.002 \pm 0.008$ & $0.015 \pm 0.002$ &  OGLE-LMC-DPV-093 & $-0.000 \pm 0.001$ & $0.028 \pm 0.006$ \\
OGLE-LMC-DPV-031 & $0.010 \pm 0.006$ & $0.293 \pm 0.005$ &  OGLE-LMC-DPV-094 & $-0.003 \pm 0.001$ & $0.047 \pm 0.004$ \\
OGLE-LMC-DPV-033 & $0.012 \pm 0.005$ & $0.238 \pm 0.004$ &  OGLE-LMC-DPV-095 & $-0.000 \pm 0.007$ & $0.006 \pm 0.002$ \\
OGLE-LMC-DPV-039 & $-0.007 \pm 0.005$ & $0.084 \pm 0.003$ & OGLE-LMC-DPV-100 & $-0.010 \pm 0.010$ & $0.114 \pm 0.012$ \\
OGLE-LMC-DPV-041 & $0.006 \pm 0.007$ & $0.028 \pm 0.004$ &  OGLE-LMC-DPV-102 & $0.014 \pm 0.008$ & $0.066 \pm 0.008$ \\
OGLE-LMC-DPV-042 & $0.007 \pm 0.004$ & $0.020 \pm 0.003$ &  OGLE-LMC-DPV-104 & $0.004 \pm 0.005$ & $0.223 \pm 0.008$ \\
OGLE-LMC-DPV-043 & $0.009 \pm 0.007$ & $0.052 \pm 0.002$ &  OGLE-LMC-DPV-106 & $0.021 \pm 0.008$ & $0.362 \pm 0.005$ \\
OGLE-LMC-DPV-044 & $0.002 \pm 0.006$ & $0.069 \pm 0.002$ &  OGLE-LMC-DPV-110 & $0.002 \pm 0.005$ & $0.150 \pm 0.003$ \\
OGLE-LMC-DPV-047 & $0.011 \pm 0.004$ & $0.053 \pm 0.003$ &  OGLE-LMC-DPV-113 & $0.021 \pm 0.007$ & $0.238 \pm 0.009$ \\
OGLE-LMC-DPV-048 & $-0.019 \pm 0.007$ & $0.207 \pm 0.002$ & OGLE-LMC-DPV-115 & $0.018 \pm 0.006$ & $0.425 \pm 0.004$ \\
OGLE-LMC-DPV-052 & $-0.021 \pm 0.008$ & $0.291 \pm 0.011$ & OGLE-LMC-DPV-116 & $0.017 \pm 0.007$ & $0.115 \pm 0.004$ \\
OGLE-LMC-DPV-054 & $0.007 \pm 0.009$ & $0.018 \pm 0.004$ &  OGLE-LMC-DPV-119 & $0.004 \pm 0.010$ & $0.250 \pm 0.011$ \\
OGLE-LMC-DPV-055 & $-0.011 \pm 0.004$ & $0.077 \pm 0.003$ & OGLE-LMC-DPV-122 & $-0.002 \pm 0.008$ & $0.192 \pm 0.011$ \\
OGLE-LMC-DPV-057 & $-0.009 \pm 0.008$ & $0.140 \pm 0.012$ & OGLE-LMC-DPV-125 & $0.003 \pm 0.005$ & $0.101 \pm 0.005$ \\
OGLE-LMC-DPV-059 & $-0.005 \pm 0.006$ & $0.067 \pm 0.009$ & OGLE-LMC-DPV-126 & $-0.003 \pm 0.002$ & $0.026 \pm 0.013$ \\
OGLE-LMC-DPV-067 & $0.015 \pm 0.008$ & $0.105 \pm 0.004$ &  OGLE-LMC-DPV-128 & $ 0.012 \pm 0.005$ & $0.053 \pm 0.006$ \\
OGLE-LMC-DPV-068 & $-0.001 \pm 0.010$ & $0.038 \pm 0.009$ & OGLE-LMC-DPV-130 & $ 0.011 \pm 0.005$ & $0.044 \pm 0.006$ \\
OGLE-LMC-DPV-069 & $0.017 \pm 0.010$ & $0.128 \pm 0.003$ &  OGLE-LMC-DPV-132 & $-0.002 \pm 0.006$ & $0.007 \pm 0.007$ \\
OGLE-LMC-DPV-072 & $-0.001 \pm 0.005$ & $0.034 \pm 0.005$ & OGLE-LMC-DPV-133 & $ 0.012 \pm 0.005$ & $0.034 \pm 0.006$
\end{longtable}
\tablefoot{$\Delta A_{II}$ is computed using Eq.~\ref{eq-amp}, and $A_{LC}$ is derived from the GLS analysis of the disentangled light curve. Systems are grouped into eclipsing and ellipsoidal classes. Within each group, objects are separated according to the $3\sigma$ criterion: $|\Delta A_{II}| > 3\sigma$ indicates significant changes around the secondary minimum, while $|\Delta A_{II}| \leq 3\sigma$ corresponds to non-significant cases.}

\end{appendix}

\end{document}